\documentclass[sigconf, screen, nonacm]{acmart}

\usepackage{color}
\usepackage{soul}    %
\usepackage{amsmath}
\usepackage{subfig}
\usepackage{xcolor, colortbl}
\usepackage{bm}
\usepackage{amsfonts}
\usepackage{acmart-taps}
\usepackage{hyperref}

\def\num#1{#1}

\usepackage{gensymb} %

\usepackage{enumitem} %

\usepackage{xspace} %

\newcommand{\systemName}{\textsc{FontCraft}\xspace}

\newcommand{\etal}{{\it{et~al.}}}
\newcommand{\ie}{i.e.,}
\newcommand{\eg}{e.g.,}

\definecolor{gray}{rgb}{0.5,0.5,0.5}
\definecolor{green}{rgb}{0, 0.6, 0}
\definecolor{orange}{rgb}{1, 0.5, 0}
\definecolor{mahogany}{rgb}{0.75, 0.25, 0.0}
\definecolor{purple}{rgb}{0.6, 0, 0.6}
\definecolor{darkgreen}{rgb}{0, 0.3, 0}
\definecolor{orange}{rgb}{1, 0.5, 0.}
\definecolor{lightblue}{rgb}{0.52, 0.75,0.91}

\newcommand{\bestcell}[1]{\cellcolor{lightblue!50}#1}

\colorlet{soullightblue}{lightblue!50}
\newcommand{\besthint}[1]{\sethlcolor{soullightblue}\hl{#1}}
\colorlet{soullightyellow}{yellow!40}

\copyrightyear{2025}
\acmYear{2025}
\setcopyright{acmlicensed}\acmConference[CHI '25]{CHI Conference on Human Factors in Computing Systems}{April 26-May 1, 2025}{Yokohama, Japan}
\acmBooktitle{CHI Conference on Human Factors in Computing Systems (CHI '25), April 26-May 1, 2025, Yokohama, Japan}
\acmDOI{10.1145/3706598.3713863}
\acmISBN{979-8-4007-1394-1/25/04}

\begin{document}

\title[\systemName: Multimodal Font Design Using Interactive Bayesian Optimization]{\systemName: Multimodal Font Design\\Using Interactive Bayesian Optimization}
\hypersetup{
  pdftitle={\systemName: Multimodal Font Design Using Interactive Bayesian Optimization},
}

\author{Yuki Tatsukawa}
\orcid{0009-0003-5128-8032}
\affiliation{%
 \institution{The University of Tokyo}
 \country{Japan}
}
\author{I-Chao Shen}
\orcid{0000-0003-4201-3793}
\affiliation{%
 \institution{The University of Tokyo}
 \country{Japan}
}
\author{Mustafa Doga Dogan}
\orcid{0000-0003-3983-1955}
\affiliation{%
 \institution{Adobe Research}
 \country{Switzerland}
}
\author{Anran Qi}
\orcid{0000-0001-7532-3451}
\affiliation{%
 \institution{Centre Inria d'Université Côte d'Azur}
 \country{France}
}
\author{Yuki Koyama}
\orcid{0000-0002-3978-1444}
\affiliation{%
 \institution{National Institute of Advanced Industrial Science and Technology (AIST)}
 \country{Japan}
}
\author{Ariel Shamir}
\orcid{0000-0003-4201-3793}
\affiliation{%
 \institution{Reichman University}
 \country{Israel}
}
\author{Takeo Igarashi}
\orcid{0000-0002-5495-6441}
\affiliation{%
 \institution{The University of Tokyo}
 \country{Japan}
}

\renewcommand{\shortauthors}{Tatsukawa, et al.}

\begin{abstract}
Creating new fonts requires a lot of human effort and professional typographic knowledge.
Despite the rapid advancements of automatic font generation models, existing methods require users to prepare pre-designed characters with target styles using font-editing software, which poses a problem for non-expert users.
To address this limitation, we propose \systemName, a system that enables font generation without relying on pre-designed characters.
Our approach integrates the exploration of a font-style latent space with human-in-the-loop preferential Bayesian optimization and multimodal references, facilitating efficient exploration and enhancing user control.
Moreover, \systemName allows users to revisit previous designs, retracting their earlier choices in the preferential Bayesian optimization process.
Once users finish editing the style of a selected character, they can propagate it to the remaining characters and further refine them as needed.
The system then generates a complete outline font in OpenType format.
We evaluated the effectiveness of \systemName through a user study comparing it to a baseline interface.
Results from both quantitative and qualitative evaluations demonstrate that \systemName enables non-expert users to design fonts efficiently.

\end{abstract}

\begin{CCSXML}
<ccs2012>
   <concept>
       <concept_id>10003120.10003121</concept_id>
       <concept_desc>Human-centered computing~Human computer interaction (HCI)</concept_desc>
       <concept_significance>300</concept_significance>
       </concept>
 </ccs2012>
\end{CCSXML}

\ccsdesc[300]{Human-centered computing~Human computer interaction (HCI)}

\keywords{font design, outline fonts, human-in-the-loop, latent space exploration, novice user support tools, generative models, typography tools}

\begin{teaserfigure}
  \centering
  \includegraphics[width=\linewidth]{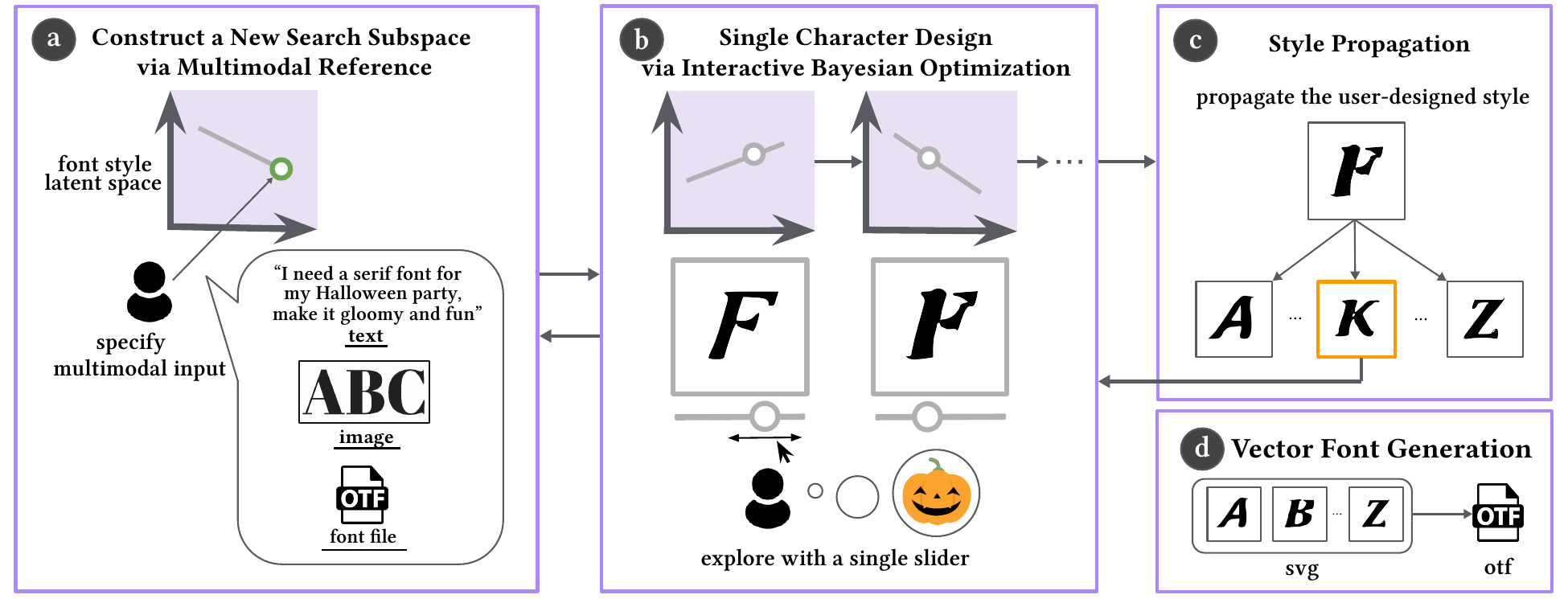}
  \caption{
  \systemName allows non-expert users to create a font without pre-designed characters through four key steps.
  (a) Users input multimodal data (text, images, font files) to construct a new search subspace.
  (b) Users repeatedly explore the search subspace recommended by Bayesian optimization or constructed by multimodal reference using a slider.
  (c) Users can propagate an edited character's style to the remaining characters and refine any unsatisfactory characters (\eg~``K'') by repeating tasks (a) and (b).
  (d) The system generates \textit{OpenType Font} (OTF) file.
  }
  \Description{(a) -- (b) --- (c)---}
  \label{fig:teaser}
\end{teaserfigure}

\maketitle

\section{Introduction}
\label{introduction}
Designing a new font for posters, websites, and advertisement banners is a challenging task, even for professional designers.
It requires a significant amount of repetitive manual effort because the designers need to create a whole set of characters.
For example, a Roman font contains \num{62} characters including ``A--Z'', ``a--z'', and ``0--9''.
Moreover, when it comes to other writing systems, for example, it takes about \num{12} months for three to five experts to design a \textit{GB18030-2000} Chinese font comprising \num{27,533} characters, according to \textit{FounderType}\footnote{\url{https://www.foundertype.com/}}, a Chinese font company.
Additionally, font design necessitates adherence to typography-specific criteria to ensure that fonts maintain consistency, meaning all characters share a uniform style.

To reduce the manual effort required for designing fonts, many previous methods have leveraged generative models to generate accurate and diverse fonts~\cite{YuchenRewrite2016, YuchenZi2zi2017, UpchurchA2Z2016, liu2023dualvector, xia2023vecfontsdf, wang2021deepvecfont, wang2023deepvecfontv2, thamizharasan2024vecfusion, JiangDCFont2017, XieDGFont2021, ZhangEMD2018, liu2024qtfont, yang2024fontdiffuser}.
These methods formulate font generation as a \emph{style transfer} problem: The style of predesigned character examples is transferred to the target characters while preserving their structure.
Although these methods generate high-quality fonts, they still have a limitation that hinders their usefulness for non-expert users in designing new fonts.
Specifically, they require users to create character examples in desired font styles using font-editing software, which poses a challenge for non-expert users.
For instance, \textit{DualVector}~\cite{liu2023dualvector}, one of the latest Roman font generation models, requires \num{3}--\num{5} predesigned character examples.

To tackle the challenges of font creation, we propose \systemName, a novel system that allows users to create new fonts \emph{\textbf{without needing to prepare specific character examples}}, making it especially user-friendly for non-experts.
As shown in~\autoref{fig:teaser}, users can create desired fonts containing numerous characters by iteratively adjusting a slider and providing multimodal references.
Our system allows them to explore within a subspace of the font style latent space of a font generative model.
The process begins with users exploring the style for a single character, such as ``A'' or ``z,'' and then propagating a selected style to the remaining characters.
Once the style is propagated, users can choose an unsatisfactory character (if any) and refine it using the same design procedure as they did with the initial character.
By repeating this process, users can ultimately design a font that meets their preferences.
Users then obtain a complete outline font in OpenType format.

The core of our system is a novel latent space exploration method that combines human-in-the-loop preferential Bayesian optimization (PBO) and multimodal references. 
Our method has two main technical contributions: \emph{\textbf{multimodal-guided subspace}} and \emph{\textbf{retractable preference modeling}}, which addresses two key limitations in existing human-in-the-loop PBO.

Human-in-the-loop PBO has been widely used to obtain the optimal solution of user preference function for visual design parameter adjustment~\cite{KoyamaSequential2017}, photographic lighting design~\cite{bayelight}, and exploring generative images and melodies~\cite{chong2021interactive,ZhouEasyGeneration2011}.
Similarly, in our system, users explore the font style latent space of a font generation model by selecting their preferred styles from candidates recommended through Bayesian optimization.
However, relying solely on PBO in the design process can diminish users' sense of agency, creativity, and ownership~\cite{Chan2022}.
To address this issue, we propose to construct multimodal-guided subspaces, which enables users to directly convey their preference to the PBO process using texts and images.
Specifically, we map user-provided multimodal references to points in the search subspace by encoding fonts that are similar to these references from an existing font database~\cite{o2014exploratory}.
To retrieve these similar fonts, our method leverages FontCLIP~\cite{tatsukawa2024fontclip}, a typography visual-language model, and constructs a new search subspace that incorporates the encoded points from the retrieved fonts.
By combining the multimodal-guided subspace with the subspace generated by the previous Bayesian optimization method~\cite{KoyamaSequential2017}, our approach enables users to design their desired fonts more efficiently.

Additionally, previous PBO methods assume that users' preferences remain consistent throughout the design process~\cite{KoyamaGallery2020}.
As a result, users cannot retract their preferences during the font design process.
To overcome this limitation, we introduce a history interface that supports retractable preference modeling.
This interface allows users to review their design history, revisit earlier states, and restart from a specified past design.
This feature is particularly valuable when users change their preferences during the design process, freeing them from the limitations of an irretractable workflow.

Furthermore, we introduce a style propagation and refinement feature, enabling users to achieve consistent styling across all characters easily.
Once users design a character with the desired style, they can propagate that style to all other characters.
They can then fine-tune any characters that require additional adjustments until satisfied.
This feature not only simplifies the design process but also ensures consistent styling throughout the entire font set.

To the best of our knowledge, \systemName is the first system that enables font design utilizing efficient font style exploration without requiring pre-designed character examples, significantly lowering the barrier to font design.
To accomplish this, our system integrates a human-in-the-loop Bayesian optimization method utilizing multimodal input with well-organized features such as history interface and style propagation.

We conducted a user study to evaluate how efficiently non-expert users could design fonts using our system, both quantitatively and qualitatively.
The study compared our system to a baseline system that solely relied on a single slider with basic PBO.
The study aimed to verify the advantages of key features in our proposed system, including \textit{the integration of PBO with \textbf{multimodal input}} and features such as \textit{history interface for retractable preference modeling} and \textit{the combination of style propagation and refinement}.
In the user study, participants without font design experience were tasked with creating fonts using both our system and the baseline.
We also collected feedback from the participants to assess their satisfaction with our system.
Both quantitative and qualitative analyses of the fonts created by participants revealed that our system produced better font designs compared to the baseline.
Survey responses also confirmed that the proposed system features significantly enhanced the efficiency of the font design process.
Additionally, we demonstrated that our system \emph{\textbf{supports other writing systems}}, such as Chinese, Japanese, and Korean (CJK), and is effective in practical applications, including logo and advertisement design.
These findings show that our system is not only applicable to non-Roman font design but also useful in practical, real-world design scenarios.

\paragraph{Contributions}
To sum up, we make the following contributions:
\begin{itemize}[leftmargin=0.5cm]
    \item We present \systemName, an interactive font design system that simplifies the process of creating fonts across various writing systems, making the design process accessible to non-experts.
    \item We propose a method that combines human-in-the-loop Bayesian optimization and multimodal references, enabling user interaction within the multimodal-guided subspace.
    \item We introduce a history interface that allows users to retract and update their preferences during the design process, which cannot be done in previous human-in-the-loop PBO methods.
    \item We develop an iterative style propagation and refinement method to ensure a consistent style throughout the font set.
\end{itemize}

\section{Related Work}

\subsection{Automatic Font Generation}
Font generation aims to create characters with a specific font style, ultimately leading to the creation of new font libraries.
Researchers have proposed various methods for generating bitmap Roman fonts, such as blending styles from template fonts~\cite{Igarashi2010}, constructing a font manifold~\cite{CampbellFontManifold2014}, and manipulating attribute scores~\cite{wang2020attribute2font}.
Additionally, recent studies have focused on synthesizing outline fonts in vector format using deep generative networks~\cite{liu2023dualvector, xia2023vecfontsdf, wang2021deepvecfont, wang2023deepvecfontv2, thamizharasan2024vecfusion}.
These works tackle the challenges by representing character outlines as sequences of tokens~\cite{wang2021deepvecfont, wang2023deepvecfontv2} or signed distance functions (SDFs)~\cite{liu2023dualvector, xia2023vecfontsdf}.
While these approaches successfully synthesize vector fonts, non-expert users may find it difficult to use them directly, as they require pre-designed characters in target fonts or manipulating various kinds of attribute scores~\cite{wang2020attribute2font}.

On the other hand, creating Chinese, Japanese, and Korean (CJK) fonts, which consist of a vast number of complex characters, requires different approaches than generating Roman fonts.
Some methods attempt to generate CJK fonts by utilizing extracted metadata, such as radicals and strokes~\cite{LianEasyFont2017, SounghuaAutomaticGeneration2009, ZhouEasyGeneration2011, ZongStrokeBank2014}.
However, these approaches face significant challenges, particularly the need for a large number of character examples.
For example, generating a font with \num{2,550} characters requires \num{522} character examples in the desired font style~\cite{LianEasyFont2017}.

To overcome these problems, recent deep learning-based works~\cite{YuchenZi2zi2017,ChaDMFont2020,JiangDCFont2017,ParkMultipleHeads2021,SunSAVAE2018} treat the font generation problem as a style transfer problem.
However, these methods require labeled data, such as radicals of characters.
In contrast, several approaches aim to train font-generation models without relying on domain knowledge~\cite{JiangDCFont2017, XieDGFont2021, ZhangEMD2018, liu2024qtfont, yang2024fontdiffuser}.
Among them, \textit{DG-Font}~\cite{XieDGFont2021} combines style and content using adaptive instance normalization (\textit{AdaIN} \cite{HuangAdaIN2017}), a straightforward yet effective style transfer technique that aligns the mean and variance of content with those of style.
This method requires only a few character examples in the desired font.
More recently, diffusion model-based methods~\cite{ho2020denoising} have achieved high-quality and high-resolution font generation~\cite{liu2024qtfont, yang2024fontdiffuser, he2024difffont, fu2024MSD}.

In this paper, we utilize the extended \textit{DG-Font} to generate fonts without the need to prepare character examples.
Although \textit{DG-Font} is not the latest model, its latent space is easier to explore than the latent space of diffusion-based font-generative models~\cite{liu2024qtfont, yang2024fontdiffuser, he2024difffont}.
Notably, our proposed system is compatible with other pretrained font generative models that utilize a font style latent space.

\subsection{Human-in-the-Loop Bayesian Optimization}
Bayesian optimization~\cite{Brochu2010B, ShahriaiReviewBO2016} is a widely used method for optimizing black-box functions.
It is particularly useful for functions that are expensive to evaluate because it aims to find the optimal value with minimal iteration, which is achieved by selecting queries that are most effective in terms of exploration and exploitation.

To reduce the number of expensive human evaluations, researchers have tried to integrate Bayesian optimization with human-in-the-loop systems~\cite{Brochu2010A, Brochu2007, KoyamaGallery2020, KoyamaSequential2017, ZhouGenerativeMelody2020, kadner2021adaptifont,Mo2024, chong2021interactive}.
For instance, Koyama~\etal\ \cite{KoyamaSequential2017} propose a method called Sequential Line Search (SLS), which finds the optimal value in the multi-dimensional space by tweaking a one-dimensional slider that is easy for humans to perform.
The line explored by the one-dimensional slider connects the point expected to be optimal and the point at which the acquisition function is maximized.
Building on SLS, Zhou~\etal~\cite{ZhouGenerativeMelody2020} propose a framework for generating melody compositions.
Their framework transforms the task of adjusting a one-dimensional slider into selecting the most favorable candidate from a set of options.
This adaptation enhanced user interaction while leveraging the strengths of Bayesian optimization.
Kadner~\etal~\cite{kadner2021adaptifont} introduce a human-in-the-loop system for font generation, focusing on optimizing fonts for readability through Bayesian optimization.
In contrast, our work expands the scope of font design by integrating SLS with multimodal references, simplifying the creation of fonts for a variety of applications beyond readability.

A notable limitation of human-in-the-loop Bayesian optimization is that it tends to reduce user agency in the design process and decrease their sense of ownership over the outcomes~\cite{Chan2022}.
Chan~\etal~\cite{Chan2022} suggest that enhancing users' ability to express their ideas to the optimizer can effectively improve both agency and ownership.
Previous works have addressed this issue by enabling users to directly incorporate preferences in various approaches, such as specify areas in the design space they wish to exclude~\cite{Mo2024}, edit the generated melody~\cite{ZhouGenerativeMelody2020} and images~\cite{chong2021interactive} directly.
While these approaches allow users to incorporate their preferences directly into the Bayesian optimization process, they assume that users' preferences are time-invariant~\cite{KoyamaGallery2020} and restrict users to a forward-directed design workflow, limiting flexibility in revisiting or re-evaluating earlier steps.

In contrast, our proposed method enables users to incorporate their preferences into the Bayesian optimization process using multimodal references, including text input.
This multimodal interactive capability is a novel improvement over previous methods.
Additionally, we provide a user interface that effectively visualizes the interaction history between the user and the system. 
This visualization allows users to easily understand the design history and revisit or re-evaluate earlier points, freeing them from the constraints of a strictly forward-directed optimization process.

\newcommand{\argmax}{\mathop{\rm arg~max}\limits}
\begin{figure*}[htb]
    \centering
    \includegraphics[width=0.95\linewidth]{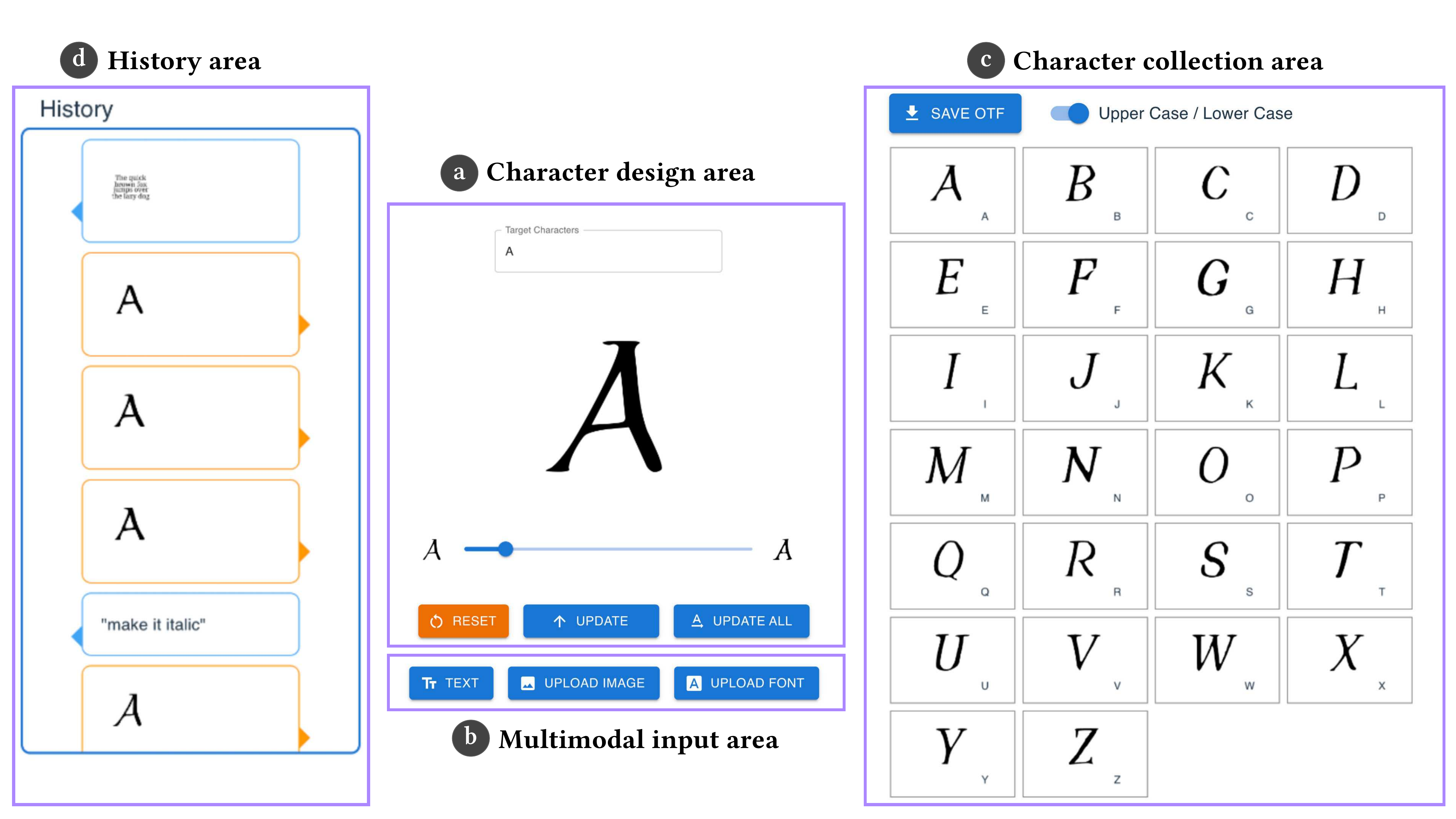}
    \caption{
    \textbf{\systemName UI.}
    Users manipulate the slider in (a) the character design area to explore the line search subspace provided by the system.
    They can also input multimodal references using (b) the multimodal input area.
    They can obtain a new recommendation by pressing the \textsc{Update} button.
    Once users are satisfied with the current style of the focused character, they can propagate its style to all other characters by pressing the \textsc{Update All} button, and the results can be viewed in (c) the character collection area.
    Optionally, users can select another character and further refine it.
    (d) The history area shows the sequence of user inputs and system outputs, enabling users to easily track their exploration history and revert to a specific checkpoint if needed.
    }
    \label{fig:UI}
\end{figure*}

\section{\systemName System Overview}

\subsection{System Architecture}
The overall architecture of \systemName, an effective system for font design, consists of two main components: the user interface (UI) and the font generative model.
Users use UI to explore the font-style latent space and select desired font styles.
The font generative model generates the bitmap representation of characters from latent variables (see \autoref{sec:fontgenerativemodel}).

\subsection{User Interface}
The UI is designed to be simple and user-friendly, particularly for users who have never designed fonts before.
As shown in \autoref{fig:UI}, our UI contains the character design area, the multimodal input area, the character collection area, and the history area.
Users interact with the system by adjusting a slider or providing multimodal data, and they can check real-time previews of generated fonts.
This interactive feedback loop allows users to iteratively refine their font choices based on visual aesthetics.
Each of these elements is crucial in facilitating the font design process.
We introduce them in this subsection.

\subsubsection{Character Design Area}
\label{sec:primaryComponent}
The character design area is the most frequently used element for exploration within the UI.
It consists of a single slider, an image viewer, and three control buttons.
The slider enables users to explore the one-dimensional latent subspace determined by either Bayesian optimization or multimodal reference inputs (see \autoref{sec:multiModalBayesianOptimization}).
The image viewer shows the currently focused character (``A'' in \autoref{fig:UI}) in the selected style, rendered in vector format.
We generate the vector character in the following steps.
First, the bitmap character is generated by the font generative model using the style latent vector selected by the handle on the slider.
Then, we reduced artifacts in the generated bitmap character image by simply setting any pixel with a grayscale value above a certain threshold to white.
Finally, we converted the filtered bitmap character into SVG format by tracing the outlines using the Potrace algorithm~\cite{selinger2003potrace}.

The three control buttons, \textsc{Reset}, \textsc{Update}, and \textsc{Update All}, serve specific functions:
\begin{itemize}
    \item \textsc{Reset}: to clear any accumulated preferences in Bayesian optimization for a focused character and reinitialize it, allowing users to start exploring the font style for that character from scratch.
    \item \textsc{Update}: to submit the selected point on the slider as the current user preference for a focused character, requesting the Bayesian optimization process to recommend a new search subspace for exploration in the next iteration.
    \item \textsc{Update All}: to propagate the style the user prefers for all characters and request the Bayesian optimization process to recommend the next search subspace.
\end{itemize}

\subsubsection{Multimodal Input Area}
\label{sec:multimodalInputComponent}
The multimodal input area allows users to provide multimodal references, including text, images, and existing font files.
These references are used to initialize or customize the font style exploration process by constructing a new search subspace.
By providing specific references, users can directly influence the system's output, making it easier for them to design desired fonts.
We explain how to encode the multimodal references into the font style latent space in \autoref{sec:multiModalBayesianOptimization}.

\subsubsection{Character Collection Area}
\label{fig:characterCollectionComponent}
The character collection area displays previews of the generated characters in vector format.
Users can zoom in on each character to inspect for any defects and select a specific character to focus on.
Once a character is selected, users can refine its font style in the character design area.

\subsubsection{History Area}
\label{fig:historyComponent}
The history area displays a sequence of user inputs and system outputs, allowing users to track their design progress.
Users can select any previous output to revert to that stage, which enables them to undo actions and restart the font design process from a specific point.
This feature allows users to retract undesired preferences and update their preferences during the design process.

\section{Method}
\subsection{Preliminary of Font Generative Model}
\label{sec:fontgenerativemodel}
\begin{figure}[ht]
    \centering
    \includegraphics[width=\linewidth]{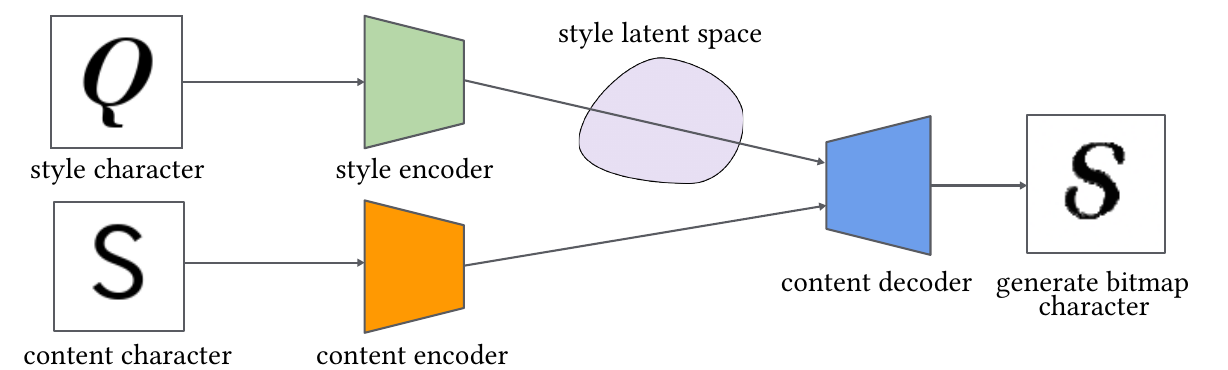}
    \caption{   
    \textbf{Overview of \textit{DG-Font}.} \textit{DG-Font} is an encoder-decoder model that takes a character image representing style and a character image representing content as input, and outputs a character image that combines the content with the specified style.
    During font designing in our system, users use our human-in-the-loop optimization to explore the style latent space of the style encoder.
    Please find the detailed architecture of the encoder and decoder in the supplemental material.
    }
    \label{fig:dg-font}
\end{figure}

We use \textit{DG-Font}~\cite{XieDGFont2021} as our font generative model. 
This model takes a style image $I_{S}$ representing the desired font style, and a content image $I_{C}$ representing the desired character, as input.
It then generates the character image that represents the desired character in the desired font style as the output. 
As illustrated in \autoref{fig:dg-font}, the architecture of \textit{DG-Font} is an encoder-decoder model with two encoders: a style encoder $E_{S}$ and a content encoder $E_{C}$, along with a content decoder $G_{C}$.
The generation process starts by extracting the style latent vector from the style image using the style encoder and the content latent vector from the content image.
Then, the content decoder takes both the style and content latent vectors as input to generate the desired font image $\hat{I}$, which maintains a similar style to the style image while preserving the structure of the content image.
Overall, the generation process during the training process can be formulated as:
\begin{align}
\bm{z}_S = E_{S}(I_S),\quad \bm{z}_C = E_{C}(I_C), \quad 
\hat{I} = G_C(\bm{z}_S, \bm{z}_C).
\end{align}
In our work, we use an enhanced version of \textit{DG-Font}, which includes an additional content discriminator.
Notably, our system (\systemName) only needs the pretrained model to generate new fonts instead of training a model from the beginning.
We will provide the details of this additional model architecture and training process in the supplemental material.

In the rest of the paper, if we need to specify the content image $I_C$ or its latent vector $\bm{z}_C$ for a specific character such as ``A'', we will denote it as  $I_C[\text{``A''}]$ $\bm{z}_C[\text{``A''}]$.
Otherwise, we will use $I_C$ or $\bm{z}_C$ for abbreviation. 
This also applied to the generated image $\hat{I}$.

\subsection{Preliminary of FontCLIP}
To incorporate multimodal input when designing fonts, we use FontCLIP~\cite{tatsukawa2024fontclip} to extract typographical features from both text and image input.
FontCLIP is a visual-language model that
bridges the semantic understanding of a large vision-language model with typographical knowledge.
It consists of a text encoder and a visual encoder and builds a joint latent space that encodes typographical knowledge.
In this joint latent space, similar font images and text prompts will have similar latent vectors.
For example, a bold font will have a similar latent vector to the text prompt ``This is a \textit{strong} and \textit{thick} font'' compared to the text prompt ``This is a \textit{thin} font''.
Therefore, the FontCLIP latent vector can be used to retrieve similar fonts using text or image input.
In our system, we utilize both the FontCLIP text encoder and visual encoder to extract a latent vector from the multimodal input to customize the linear subspace.

\subsection{Preliminary of Human-in-the-Loop Bayesian Optimization}
\label{method:prelim_BO}
\subsubsection{Problem Formulation}
Human-in-the-loop optimization is a computational approach used to solve parameter optimization problems involving human evaluators in its iterative algorithm.
It is commonly used to support design tasks that involve generating visual content defined by a set of parameters $\bm{x}$, with the aim of achieving certain subjective design goals.
Specifically, we can formulate such an optimization problem as:
\begin{align}
\bm{x}^* = \argmax_{\bm{x}\in{\mathcal{X}}}f(\bm{x}),
\label{eq:opt_goal}
\end{align}
where $\mathcal{X}$ is the search space, and $f: \mathcal{X} \rightarrow \mathbb{R}$ is the goodness function to evaluate a subjective design goal (\eg~the aesthetics of the current design).
We aim to find the optimal value $\bm{x}^*$ with the fewest trials because evaluating $f(\cdot)$ is costly.
However, solving \autoref{eq:opt_goal} using traditional Bayesian optimization (BO) might not be suitable for many design tasks.
This is because it is often difficult to assign an exact value to a sample, whereas comparing a couple of samples and choosing the preferred one is more intuitive.
For example, it is hard for users to give a score to a font individually, but easier for them to choose the preferred font from a set of candidates.
Therefore, in this work, we choose to use preferential Bayesian optimization (PBO) \cite{Koyama2022}, which is a variant of Bayesian optimization (BO) that runs with relative preferential data.
In particular, we build our method upon Sequential Line Search (SLS) \cite{KoyamaSequential2017}, a PBO method that constructs a sequence of linear subspaces that leads to the optimal parameters that match the user's need.

\subsubsection{Sequential Line Search (SLS)}
\label{sec:sls}
With SLS, the user can search for his/her preference by adjusting a slider.
At each iteration of the optimization, SLS constructs a linear subspace using the current-best position $\bm{x}^{+}$ and the best-expected-improving position $\bm{x}^{\text{EI}}$.
Suppose we already have $t$ observed response so far, then the next linear subspace $\mathcal{S}_{t+1}$ is constructing by connecting:
\begin{align}
    \bm{x}^\text{EI}_t &= \argmax_{\bm{x}\in{\mathcal{X}}}a^\text{EI}(\bm{x};\mathcal{D}_t) \label{eq:build_subspace} \\ \nonumber
    \bm{x}^+_t &= \argmax_{\bm{x}\in{\left\{\bm{x}_i\right\}_{i=1}^{N_t}}}\mu_t(\bm{x})
\end{align}
where $\left\{\bm{x}_i\right\}_{i=1}^{N_t}$ denotes the set of points observed so far, $\mu_t$ and $a^\text{EI}$ are the predicted mean function and the acquisition function calculated from the current data.
We use the expected improvement (EI) criterion as the acquisition function to choose the next sampling point that is likely to optimize the function $f$ and at the same time its evaluation is more informative:
\begin{align} \label{equation:acquisition}
    a^{\text{EI}}(\bm{x} ; \mathcal{D})=\mathbb{E}\left[\max \left\{f(\bm{x})-f^{+}, 0\right\}\right].
\end{align}

After the $t$-th iteration, we suppose that we obtained a set of $t$ single slider responses, which is represented as
\begin{align}
    \mathcal{D}_t=\left\{\bm{x}_i^\text{chosen}>\left\{\bm{x}_{i-1}^{+}, \bm{x}_{i-1}^{\text{EI}}\right\}\right\}_{i=1}^t,
\label{eq:response_data}
\end{align}
where $\bm{x}_i^{\text{chosen}}$ represent the position chosen by the user at the $t$-th iteration.

Let $f_i$ be the goodness function value at a data point $\bm{x}_i$, i.e., $f_i = f(\bm{x}_i)$, and $\bm{f}$ be the concatenation of the goodness values of all data points:
\begin{align}
\bm{f} = [f_1, f_2, \ldots, f_N].
\end{align}
Under the assumption of Gaussian process (GP) prior on $f$, we use $\bm{\theta}$ to represent the hyperparameters of the multivariate Gaussian distribution.
Since the goodness values $\bm{f}$ and the hyperparameters $\bm{\theta}$ are correlated, we infer $\bm{f}$ and $\bm{\theta}$ jointly by using MAP estimation:
\begin{align}
(\bm{f}^{\text{MAP}}, \bm{\theta}^{\text{MAP}})&=\argmax_{(\bm{f},\bm{\theta})}p(\bm{f}, \bm{\theta} \mid \mathcal{D}) \nonumber \\
&= \argmax_{(\bm{f},\bm{\theta})}p(\mathcal{D} \mid \bm{f}, \bm{\theta}) p(\bm{f}\mid \bm{\theta}) p(\bm{\theta}).
\label{eq:map}
\end{align}

Once $\bm{f}^{\text{MAP}}$ and $\bm{\theta}^{\text{MAP}}$ have been estimated, we can compute $\mu(\cdot)$, $\sigma(\cdot)$, and $a^{\text{EI}}(\cdot)$ in order to construct the next slider subspace $\mathcal{S}_{t+1}$.
We only describe the minimum concept of how SLS constructs the linear subspace to optimize the function $f$ for understanding how we incorporate it in the style latent space of a font generative model and multimodal input.
For more details, please refer to the supplemental material.

\subsection{Multimodal Bayesian Optimization for Font Generation}
\label{sec:multiModalBayesianOptimization}
Specifically, following \autoref{eq:opt_goal}, the objective function for designing a character can be formulated as:
\begin{align}
\bm{z}^* = \argmax_{\bm{z}\in{\mathcal{Z}}}f(G(\bm{z})), \label{eq:argmax_latent_space}
\end{align}
where $\bm{z} \in\mathcal{Z}$ is the style latent vector of the font generative model, which is the search space $\mathcal{X}$ in our problem setting.
Moreover, $G$ is the decoder of the font generative model, and $f: \mathcal{Z} \rightarrow \mathbb{R}$ is the user preference function that measures how good the currently generated character is perceived by the user.
To solve \autoref{eq:argmax_latent_space} and obtain the desired font, users can perform three different actions: \textit{explore the font style latent space}, \textit{retract previous preferences}, and \textit{propagate style to other characters} at each iteration.

\subsubsection{Action 1: Explore the Font Style Latent Space}
\label{sec:explorationWithSingleSlider}
\begin{figure}[t]
    \centering
    \includegraphics[width=\linewidth]{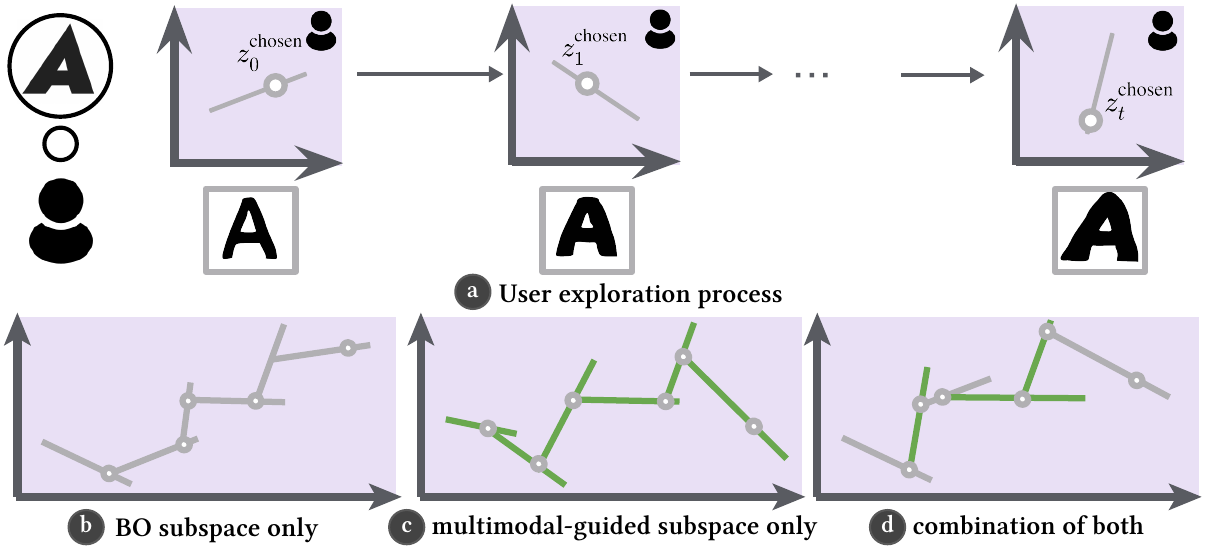}
    \caption{
    \textbf{Exploration of the font style latent space using a single slider.} 
    (a) Users explore a one-dimensional search subspace within the font style latent space using a single slider.
    At each iteration, users choose a point in the latent subspace and submit it as their current preference $\bm{z}^{\text{chosen}}_t$.
    After a couple of iterations, users gradually converge to their desired font style.
    The overall exploration process, users can explore (b) BO subspace only, (c) multimodal-guided subspace only, and (d) combination of both. 
}
    \label{fig:ExploreWithBayesianOptimization}
\end{figure}

The primary task for users is to explore the one-dimensional font-style search subspace using a slider.
By repeating the slide manipulation, users gradually converge on a point that aligns with their desired font style as illustrated in \autoref{fig:ExploreWithBayesianOptimization}.
This subspace is constructed by the system in two ways:
one solely follows the SLS method and another utilizes multimodal references.
Note that, regardless of these two different approaches to constructing the subspace, the interaction (\ie~manipulating the slider and submitting a preferred point to the system) remains consistent.

\paragraph{Constructing a SLS subspace}
Using the SLS method outlined in \autoref{sec:sls}, our system constructs a linear subspace $\mathcal{S}_{t}$ by connecting the current best point $(\bm{z}^+_t)$ and the point that maximizes the acquisition function $(\bm{z}^\text{EI}_t)$ using \autoref{eq:build_subspace} at the $t$-th iteration.
Then, users can choose a style latent vector $\bm{z}$ using the slider within $\mathcal{S}_{t}$ and view the generated character $G(\bm{z})$.
Once satisfied, users submit their preferred point on the slider $\bm{z}^{\text{chosen}}_t$ by clicking the \textsc{Update} or \textsc{Update all} button and request the system to construct a new linear subspace $\mathcal{S}_{t+1}$ for the $(t+1)$-th iteration using \autoref{eq:map}.

\paragraph{Constructing a multimodal-guided subspace}
While exploration with a single slider is useful, the linear subspace determined solely by Bayesian optimization sometimes fails to capture user preferences effectively, which leads to an increasing number of iterations and potentially causes frustration and a diminished sense of agency.
To address this issue, we allow users to intervene in the linear subspace construction by providing multimodal references at any iteration.
At $(t+1)$-th iteration, instead of exploring the linear subspace $\mathcal{S}_{t+1}=(\bm{z}^+_t, \bm{z}^\text{EI}_t)$ constructed solely by Bayesian optimization, the user explores the multimodal-guided subspace: $\mathcal{S}^{mm}_{t+1} = (\bm{z}^+_t, \bm{z}^{mm}_t)$.
Here, $\bm{z}^{mm}_t$ is the style latent vector obtained by retrieving the most similar fonts to the multimodal reference provided at $(t+1)$-th iteration from a font database containing \num{1,169} Roman fonts collected by O'Donovan~\etal~\cite{o2014exploratory}.
Specifically, we retrieve $n$ fonts and use the mean of their latent vectors as $\bm{z}^{mm}_t$ (we use $n = 5$ in our implementation).
Once the user is satisfied with the current generated character, the slider response: ($\bm{z}_{t+1}^\text{chosen},\bm{z}_{t}^{+}, \bm{z}_{t}^{mm}$) will be recorded in $\mathcal{D}_{t+1}$ (\autoref{eq:response_data}) and used for constructing the subspace in the future iteration.
This means that all multimodal references provided until iteration $t$ will affect the subspace constructed at $(t+1)$-th iteration.
In our current implementation, at each iteration, users can provide only a single multimodal reference, and we construct a new search subspace by connecting the current chosen point and the latent vector of the provided multimodal reference.
At $0$-th iteration, we construct the initial linear subspace $\mathcal{S}_{0}=(\bm{z}^{\text{init}}, \bm{z}^{mm}_{0})$, where $\bm{z}^{\text{init}}$ represents the style latent vector of a commonly used font (we use ``IPAex gothic'' font in our implementation).
Note that the multimodal references are only used to construct the linear subspace for the user to explore, not being directly used to infer the user preferences.r

To construct a multimodal subspace with the user-provided multimodal reference, our system identifies a suitable font in our font database that corresponds to the input text or image. 
For text input, the system first extracts font attributes such as ``formal,'' ``italic,'' and ``happy'' using a Large Language Model (LLM). 
The LLM selects relevant font attributes based on the given text.
We utilize $37$ types of font attributes compiled by O'Donovan~\etal~\cite{o2014exploratory} (see the supplemental material for more details).
Once the font attributes are extracted, the system obtains the feature vector of these text attributes using the text encoder of FontCLIP~\cite{tatsukawa2024fontclip} and retrieves fonts from our font database based on feature similarity. 
The retrieved fonts are used as the most suitable options, and the mean of their style latent vectors $\bm{z}^{mm}$ is obtained using the style encoder $E_{S}$.

\begin{figure*}[ht]
    \centering
    \includegraphics[width=0.95\linewidth]{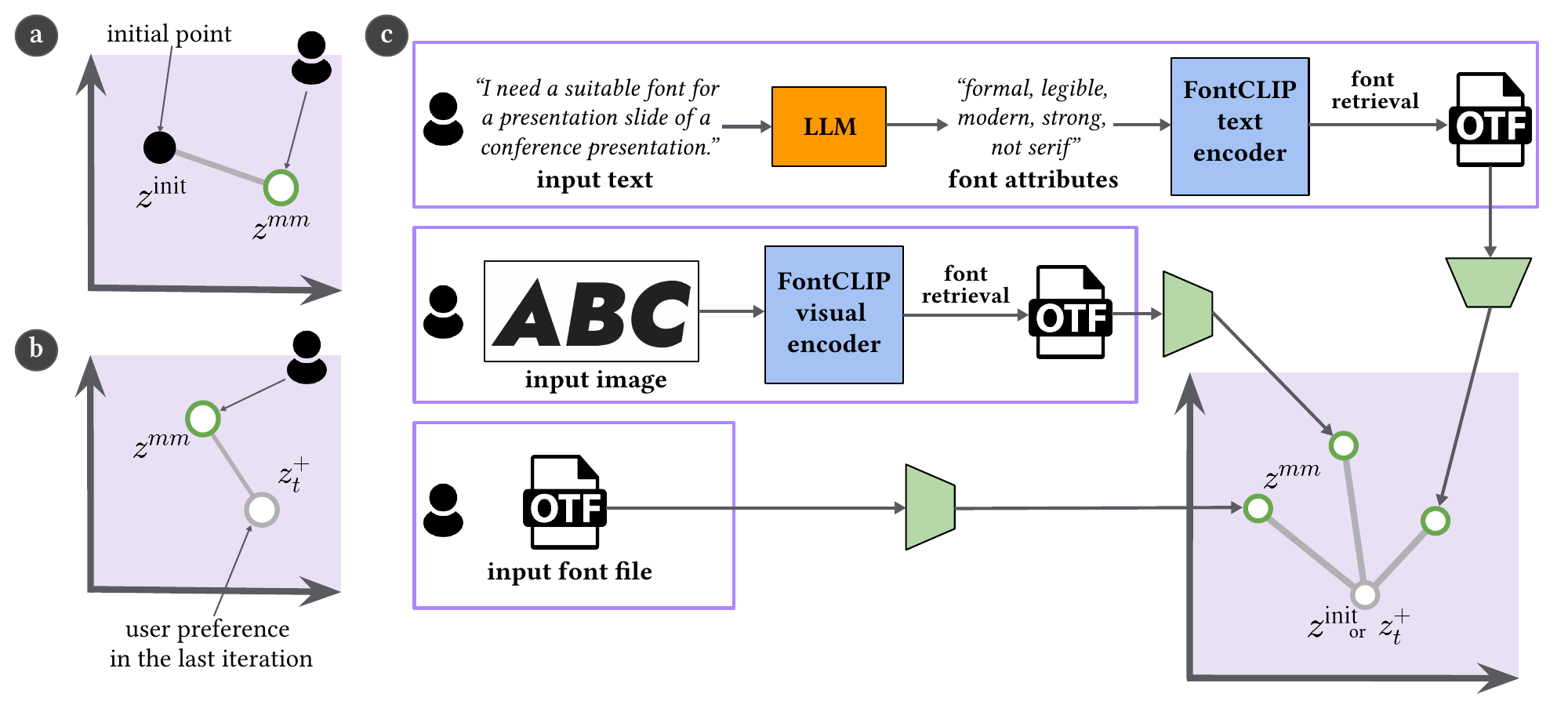}
    \caption{
    \textbf{Constructing linear subspaces using multimodal references.} 
    (a) At the start of the font design process using our proposed method, the user inputs text, an image, or a font file. 
    The system encodes this input into a font style latent vector and initializes the line search space by connecting the latent vector and a fixed point predetermined by the system.
    (b) Additionally, the user can introduce multimodal inputs at any stage of the design process.
    When the user provides new input, the system generates a new line search subspace by connecting the last user preference point with the newly encoded point.
    (c) Our system encodes multimodal input into the style latent space by leveraging LLM and FontCLIP text and visual encoders.
}
    \label{fig:multi_modal_subspace}
\end{figure*}

For the text-rendered image reference, our system retrieves its most similar font in the font database using the FontCLIP feature.
Similarly, we then project the retrieved font image into the style latent space using $E_{S}$ and obtain $\bm{z}^{mm}$.
Finally, for font file input, our system directly uses the provided font as the suitable font and projects it into the latent space, similar to the process for text and image inputs.
\autoref{fig:multi_modal_subspace}(c) illustrates how to obtain the style latent vector of the multimodal references.

\subsubsection{Action 2: Retract Previous Preferences}
At each iteration, if the user is not satisfied with the current design and the recommended candidates, they can choose to retract previous preferences.
Specifically, at $(t+1)$-th iteration, if the user wishes to retract the last two slider manipulations, then the last two slider responses stored in $\mathcal{D}_t$ will be discarded.
Then, if the user opts to explore a new subspace by providing a new multimodal reference, they will then explore the subspace $\mathcal{S}^{mm}_{t-1} = (\bm{z}^+_{t-2}, \bm{z}^{mm}_{t-2})$.
Otherwise, the user will explore a subspace constructed using SLS solely: $\mathcal{S}_{t-1} = (\bm{z}^+_{t-2}, \bm{z}^{\text{EI}}_{t-2})$.
By retracting previous preferences, users can update their preferences during the design process.

\subsubsection{Action 3: Propagate Style to Other Characters}
Once the user is satisfied with the design of the focused character (\eg~``A'') and obtains the style latent vector $\bm{z}_S^{*}$, our system can propagate the style to all other characters and finish designing a single character.
Specifically, to propagate the style vector to character ``B'':
\begin{align}
\hat{I}[\text{``B''}] = G_C(\bm{z}_S^{*}[\text{``A''}], \bm{z}_C[\text{``B''}]).
\end{align}
Next, users can check all characters with the propagated styles.
If they are unsatisfied with the result of another character, they can perform action 1 or action 2 to design that character.
If they are satisfied, the resulting font is exported as an outline font.

\section{Evaluation}
\label{evaluation}

\subsection{Simulated Evaluation of Multimodal Reference}
\label{sec:simulatedEvaluation}
To quantitatively evaluate how multimodal reference can help our human-in-the-loop optimization, we designed a simulation test to compare two linear subspace initialization methods: using multimodal reference and random fonts.
\subsubsection{Procedure}
\begin{figure}[ht]
    \centering
    \includegraphics[width=\linewidth]{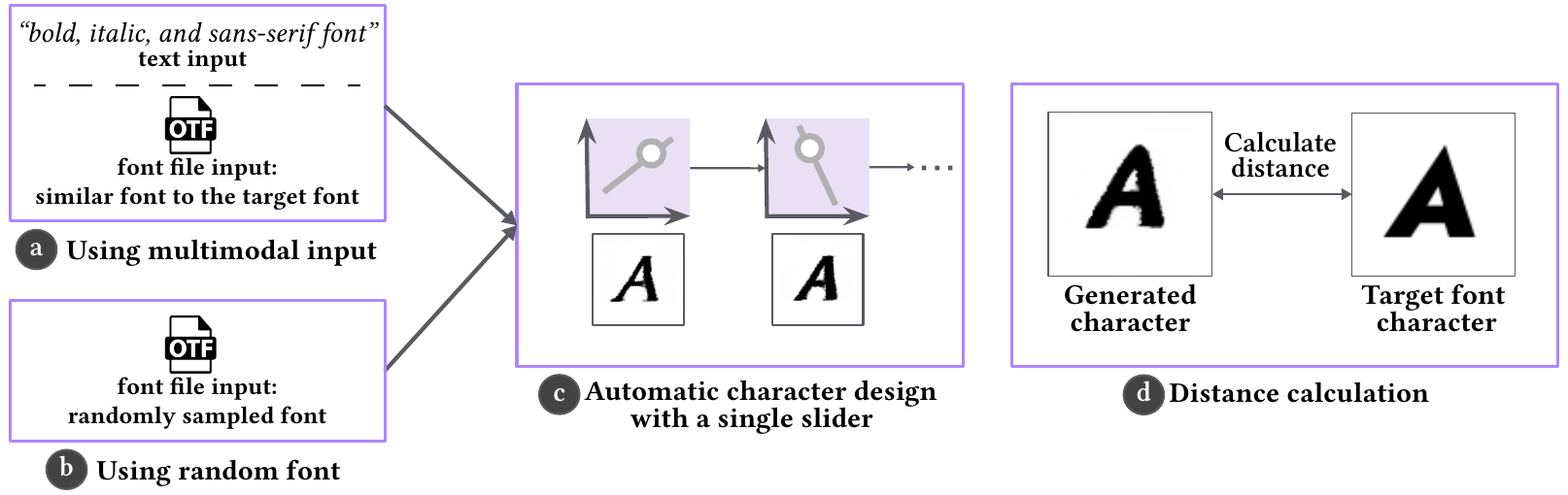}
    \caption{
    \textbf{Evaluation of linear subspace initialization methods.}
    We compared two initialization methods for exploration with Bayesian optimization.
    (a) One method uses input text or a similar font file for initialization, while (b) the other initialize method uses a randomly sampled font from a font database.
    After initialization, both methods follow the same automatic exploration process (c), where the optimal point on the single linear subspace is repeatedly identified and submitted to the system.
    In each iteration, we measure the distance between the generated character and the target font character to identify the optimal point, as shown in (d).
    Note that we use the bitmap format of the character for distance calculation, without vectorizing it.
    }
    \label{fig:multimodalInputIniitalizationEvaluation}
\end{figure}

We illustrate the procedure of the simulation test in \autoref{fig:multimodalInputIniitalizationEvaluation}. 
Given a base font character (\eg~``A''), the goal of the simulation test is to resemble the target font character (\autoref{fig:multimodalInputIniitalizationEvaluation}(d)) by exploring the style latent space through optimization.
Specifically, we simulate user selections using the following process. 
At each iteration, our method selects a point in the slider's search subspace with the minimum perceptual metric (we use \textit{DreamSim}~\cite{fu2023dreamsim}) against the target font character.
Then, the selected point is used to request Bayesian optimization to recommend the next linear subspace.
We iterate this process to observe the convergence of the optimization progress using both initialization methods.

For initializing using multimodal reference, we test \textit{text input} and \textit{font file input} in this experiment.
For the text input, we create a descriptive text that characterizes the target font and use it to initialize the search subspace.
For font file input, we manually select a font from candidate fonts that closely resembles the target font and use it to initialize the search subspace.
Finally, for the baseline method, we choose a font randomly from our font database and use it to construct the initial search subspace.

We conducted this experiment using the character ``A'' for $10$ different target fonts, randomly selected from our font database.
We collected $12$ kinds of fonts from which we chose a similar font to each target font for the font file input.
For each target font, we choose the most similar font out of the $12$ candidate fonts.
The $12$ candidate fonts consist of two popular font families, \textit{Roboto} and \textit{NotoSerif}, and each font family has six variations: \textit{Light}, \textit{Light Italic}, \textit{Regular}, \textit{Regular Italic}, \textit{Bold}, and \textit{Bold Italic}.
This selection simulated a scenario where users start with popular fonts and design new fonts based on one of these similar candidates.
For each target font, the optimization process includes $10$ iterations of Bayesian optimization.

\subsubsection{Results}
\begin{figure}[ht]
    \centering
    \includegraphics[width=0.5\textwidth]{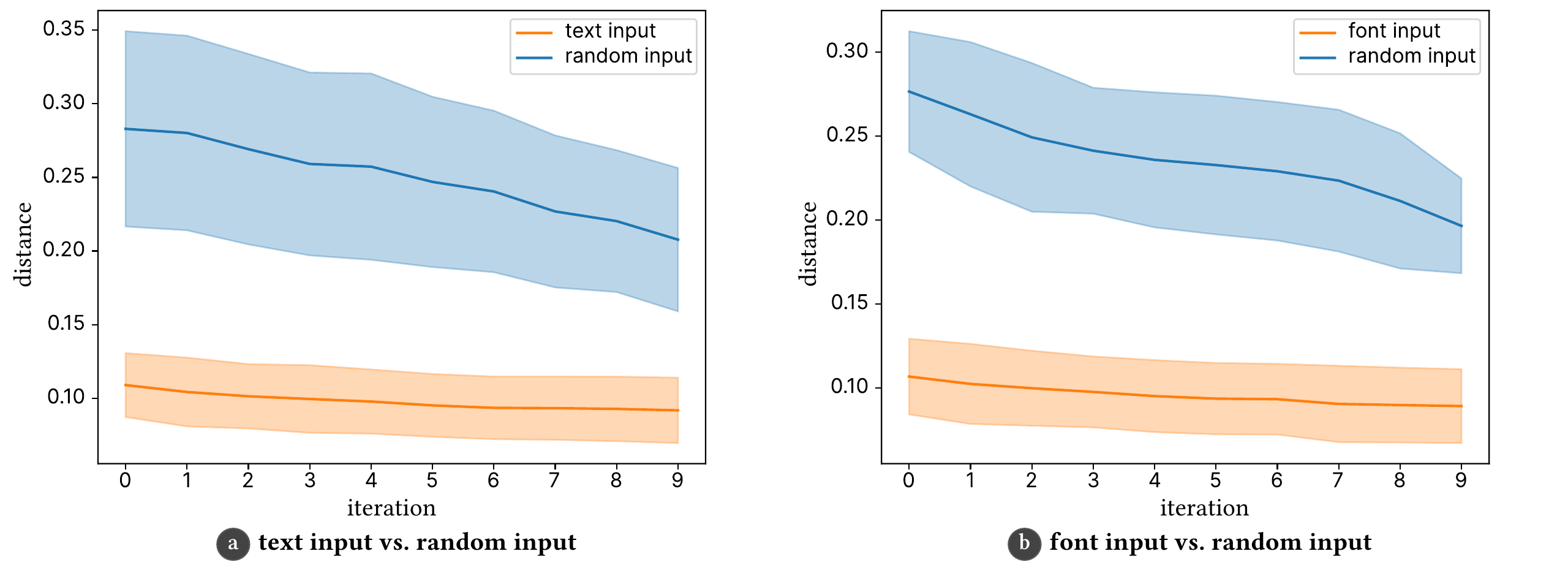}
    \caption{
    \textbf{Convergence comparison between two initialization methods.}
    The figure illustrates how the \textit{DreamSim} distance between the designed font character and the target font character converges during exploration with human-in-the-loop optimization.
    The optimization processes initialized by text font references (orange) obtain better results compared to processes initialized by random font (blue).
    }
    \label{fig:multimodalInputIniitalizationEvaluationResult}
\end{figure}

In \autoref{fig:multimodalInputIniitalizationEvaluationResult}, we show the mean and standard deviation of the distances between the optimized results and all target fonts.
We can observe that the optimization processes with text and font file references converge to a lower \textit{DreamSim} distance to the target font character compared to those initialized with a randomly selected font.
These results indicate that using multimodal references for initializing the human-in-the-loop optimization leads to more effective exploration than random initialization.

\subsection{User Study}
\label{sec:user-study}
To evaluate the effectiveness of our proposed system, we conducted a user study in which participants were asked to design fonts using both a baseline system and our system. 
The goals of this study were threefold: 
\begin{itemize}
    \item to assess the overall effectiveness of our system, including the integration of Bayesian optimization, multimodal reference, history interface, and style propagation.
    \item to compare the fonts designed by participants both qualitatively and quantitatively against those created using the baseline system.
    \item to gather qualitative feedback on the user experience with our system.
\end{itemize}

\subsubsection{Comparison Systems}
For the user study, we added a special feature called \textsc{Font Palette} to our proposed system.
By clicking the \textsc{Font Palette} button, users can view a visualization of the \num{12} popular fonts described in \autoref{sec:simulatedEvaluation} and select one to input as their preference, simplifying the process of inputting a font file.
Additionally, we removed the \textsc{Upload Image} and \textsc{Upload Font} buttons from the UI in \autoref{fig:UI} for simplicity.
As a result, users can now easily input text and font files using the \textsc{Text} and \textsc{Font Palette} buttons, respectively.

\begin{figure}[ht]
    \centering
    \includegraphics[width=0.5\textwidth]{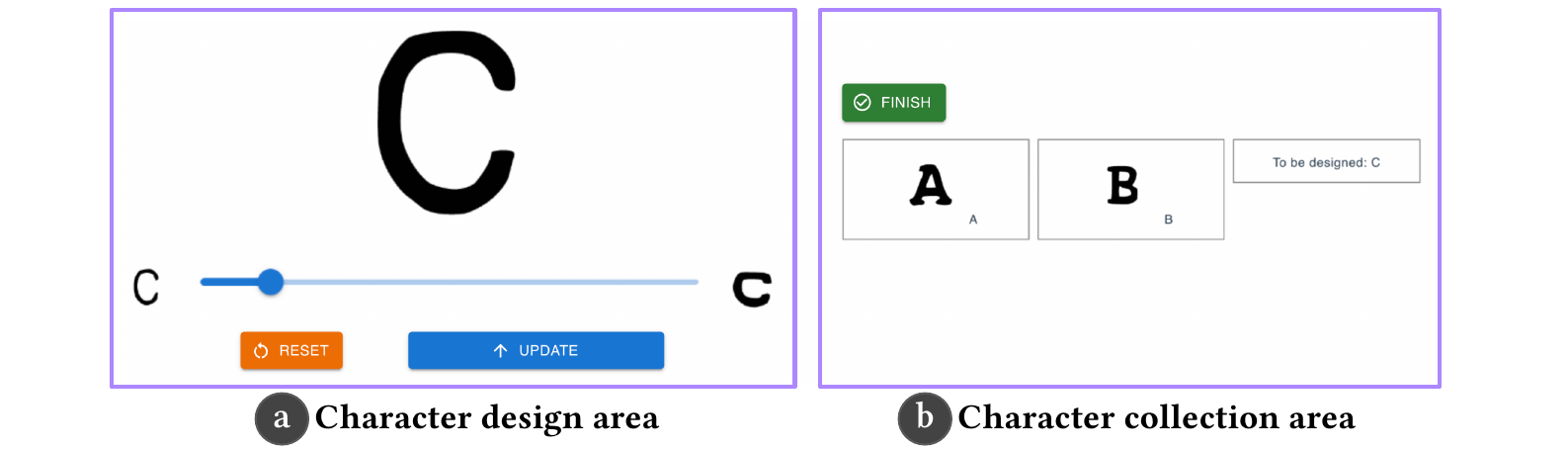}
    \caption{
    \textbf{User interface of the baseline system.}
    In the (a) character design area, users use a slider to explore the one-dimensional subspace within the font style latent space recommended by Bayesian optimization.
    By clicking the \textsc{Reset} button, users can reset their preference history in the Bayesian optimization process, randomly reinitializing the search subspace.
    The users can check the characters that they have already designed are displayed in the (b) character collection area.
}
\label{fig:baselineSystemUI}
\end{figure}

To assess the effectiveness of the multimodal reference and style propagation features in our system, we created a baseline system that includes only a single slider, as illustrated in \autoref{fig:baselineSystemUI}.
In this baseline system, users can explore the font style latent space solely by adjusting the slider, guided by the Bayesian optimization process. 
Unlike our proposed system, the baseline’s one-dimensional search space is initialized by connecting a fixed point with a randomly initialized point.
The fixed point corresponds to the style of the \textit{IPAex Gothic} font, as described in \autoref{sec:simulatedEvaluation}. 
If users encounter difficulties during exploration, they can reset their preference history in the Bayesian optimization process and restart from a newly randomized search subspace.
Additionally, this baseline method lacks a style propagation function, requiring users to design each character individually.

\subsubsection{Procedure}
We recruited ten people for the user study.
Each participant was presented with a target font and asked to design three characters, ``A'', ``B'', and ``C'' that closely match the target font using both \systemName and the baseline system.
For this user study, we prepared two target fonts, Font 1 and Font 2.
Each font design session continued until one of the following conditions was met: (1) the participant was satisfied with the quality of the characters they designed, (2) they felt that further improvement was difficult, or (3) the $7$-minute time limit was reached.
The user study followed this sequence: (Tutorial of \systemName $\rightarrow$ Font 1 with \systemName $\rightarrow$ Font 2 with \systemName $\rightarrow$ Tutorial of baseline $\rightarrow$ Font 1 with baseline $\rightarrow$ Font 2 with baseline $\rightarrow$ Survey).
The order of using \systemName and the baseline system was randomized for each participant.
After the font design sessions, participants were asked to complete a questionnaire that validated our system.
The entire user study took approximately $60$ minutes, with each tutorial lasting $10$ minutes, each font design session $7$ minutes, and the survey $10$ minutes.

\subsubsection{Results and Discussion}
\begin{figure}[ht]
    \centering
    \includegraphics[width=\linewidth]{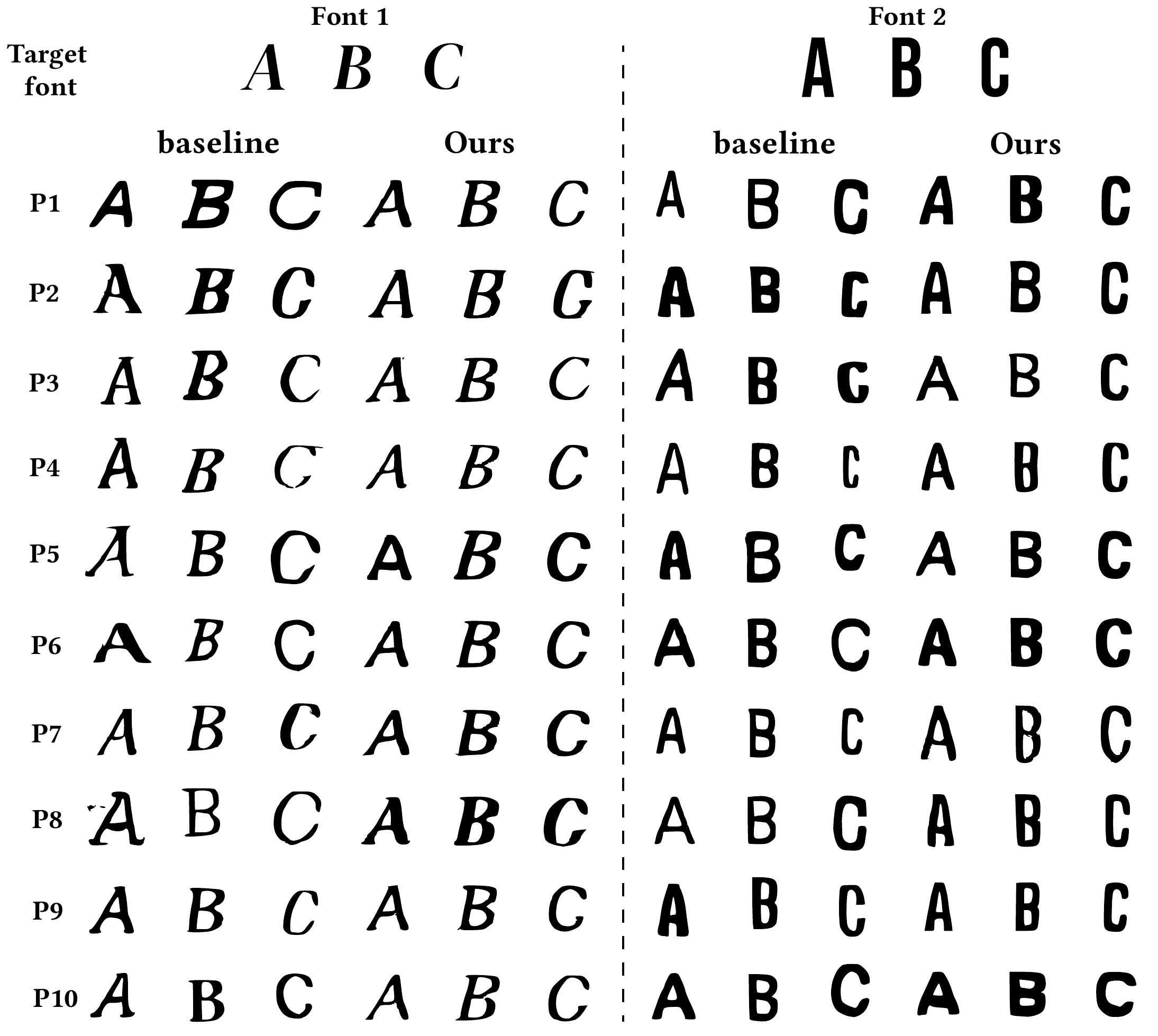}
    \caption{
    \textbf{Characters designed by user study participants.}
    Our system enables users to design characters that are more similar to the target font characters and maintain higher consistency between each other.
    In the case of Font 1, participants successfully designed all characters with the slant style using our system, while some participants failed to create the slant style for ``A'' using the baseline system.
    For Font 2, all participants designed characters with consistent styles using our system, whereas the styles of characters designed using the baseline system were inconsistent.
}
\label{fig:userStudyResult}
\end{figure}

We compared the designed fonts using our system and the baseline system both quantitatively and qualitatively.
For the quantitative evaluation, we calculated the distance between the target font characters and the designed characters in the \textit{DreamSim} latent space.
As shown in \autoref{tab:userStudyResult}(a), the characters designed with our system closely resembled the target font characters compared to those designed with the baseline system.
Additionally, we measured the style consistency between all characters designed by each participant by calculating the mean distance between the characters ``A'', ``B'', and ``C'' in the \textit{DreamSim} latent space.
As shown in \autoref{tab:userStudyResult}(b), the distance is smaller when using our system to the baseline system, indicating our system enables more style-consistent character design.
In \autoref{fig:userStudyResult}, we showed the characters designed by all participants (P1--P10).
For Font 1, the ``A'' characters designed by P2, P3, P4, P6, and P8 using our system closely matched the slanted style of the target ``A,'' while they failed to design the slanted style using the baseline system, indicating that participants effectively captured the italic feature through multimodal reference.
On the other hand, characters designed by P1, P2, P3, P4, P5, P8, and P10 using the baseline system showed inconsistencies in style within the same font (\eg~variations in size, height, and weight).
In contrast, characters designed with our system exhibited greater consistency, suggesting that style propagation helped create more cohesive designs.

\aptLtoX[graphic=no,type=html]{\begin{table}[ht]
\centering
\begin{tabular}{lll}
\multicolumn{3}{c}{\bf (a) Target font similarity $\downarrow$~~~~~}\\
\hline
                & Font 1          & Font 2          \\ \hline
Baseline & 0.1680          & 0.1416          \\
\systemName  & \bestcell{0.1591} & \bestcell{0.1355} \\ \hline
\end{tabular}
\begin{tabular}{lll}
\multicolumn{3}{c}{~~~~~\bf (b) Designed character consistency $\downarrow$}\\
\hline
                & Font 1          & Font 2          \\ \hline
Baseline & 0.3303          & 0.2893          \\
\systemName     & \bestcell{0.2983} & \bestcell{0.2793} \\ \hline
\end{tabular}
\caption{
(a) 
We calculated the distance between the characters designed by the participants and the target font characters.
Each value represents the mean distance across the $12$ characters (``A'', ``B'', ``C'' designed by the four participants).
The characters designed using our system are closer to the ground truth compared to those with the baseline system. 
(b)
We measured the character consistency between the characters ``A'', ``B'', and ``C'' designed by each participant.
Each value represents the mean distance across the three characters designed by each participant.
The distance among the three characters designed using our system is smaller than that with the baseline system, which indicates our system enables more style-consistent character design. 
($\downarrow$ denotes the lower values are better and we highlight the \besthint{best} result for each target font.)
}
\label{tab:userStudyResult}
\end{table}}{\begin{table}[ht]
\centering
\subfloat[Target font similarity $\downarrow$]{
\begin{tabular}{lll}
\toprule
                & Font 1          & Font 2          \\ \midrule
Baseline & 0.1680          & 0.1416          \\
\systemName  & \bestcell{0.1591} & \bestcell{0.1355} \\ \bottomrule
\end{tabular}
}
\subfloat[Designed character consistency $\downarrow$]{
\begin{tabular}{lll}
\toprule
                & Font 1          & Font 2          \\ \midrule
Baseline & 0.3303          & 0.2893          \\
\systemName     & \bestcell{0.2983} & \bestcell{0.2793} \\ \bottomrule
\end{tabular}
}
\caption{
(a) 
We calculated the distance between the characters designed by the participants and the target font characters.
Each value represents the mean distance across the $12$ characters (``A'', ``B'', ``C'' designed by the four participants).
The characters designed using our system are closer to the ground truth compared to those with the baseline system. 
(b)
We measured the character consistency between the characters ``A'', ``B'', and ``C'' designed by each participant.
Each value represents the mean distance across the three characters designed by each participant.
The distance among the three characters designed using our system is smaller than that with the baseline system, which indicates our system enables more style-consistent character design. 
($\downarrow$ denotes the lower values are better and we highlight the \besthint{best} result for each target font.)
}
\label{tab:userStudyResult}
\end{table}}

Next, we evaluated participant feedback to validate the effectiveness of our system.
We asked questions about the functions in our system, including slider operation, multimodal reference, style propagation, and history interface.
When we asked the question \textit{``Were you satisfied with the designed characters?''}, seven out of the ten participants answered yes, while P2 and P10 commented neutral, and P9 expressed no.
P9 noted that he observed distortions in the generated characters and felt the system was not good at generating straight lines.
In response to the question \textit{``Do you think you were able to design fonts easily with the system?''}, all ten participants answered yes, demonstrating the system's effectiveness in enabling non-expert users to design fonts with ease.

In response to the question \textit{``Do you think you were able to effectively use the slider operation for font design?''}, nine participants answered yes.
P4, who answered no, expressed dissatisfaction, stating that while the combination of slider manipulation and multimodal reference was effective, using only the slider and repeatedly clicking the \textsc{Update} button sometimes resulted in a linear subspace that excluded the desired character style.
P4 emphasized the importance of using multimodal reference at the right moments to avoid unsatisfactory suggestions and stated that relying solely on the slider was not effective.
P4 also highlighted that the history interface was useful for reverting to a previous point, leading to the escape of an undesirable search subspace suggested by the system.
P4's feedback reflects the findings suggested in Chan~\etal~\cite{Chan2022}, which indicate that designers working with BO may experience a loss of agency.
In contrast, our method provides users with a way to contribute concrete ideas that guide the BO process, thereby helping them regain a sense of agency.

In response to the question, ``\textit{Do you think you were able to effectively use text input?}'', eight of ten participants answered yes.
P1, P2, P4, P6, P7, P9, and P10 found text input helpful for making broad changes, such as adjusting weight or slant, but not for fine-tuning details or specifying complicated characteristics.
Additionally, P1, P6, and P7 mentioned that understanding typographical terms like ``bold'' and ``italic'' was necessary.
This feedback indicates that while text input is useful for exploring rough font styles, it has limitations in designing font details and requires some typographic knowledge.

Regarding the question, \textit{``Do you think you were able to effectively use the similar fonts provided by font palette?''}, eight of ten answered yes.
P4 and P7 commented that the font palette is particularly helpful when it is difficult to describe the desired font style in texts.
P8, P9, and P10 stated that initializing the search subspace using the font palette function allowed them to begin the design task more smoothly compared to the baseline system.
However, P5 expressed dissatisfaction, stating that the style of the character generated did not perfectly align with the font they selected from the font palette.
This discrepancy, caused by the encoding-decoding process of the font generative model, could lead to confusion among users.
To address this issue, it is important to communicate to users that the generated characters may not always perfectly match the multimodal reference.
Additionally, we anticipate that newer font generative models could help mitigate this discrepancy.
It is worth emphasizing that our proposed system is compatible with any font generative model, provided an efficient font style latent space can be established within it.
On the other hand, P2 explained that he did not use the font palette because he preferred to describe the target font style using text input.
This feedback suggests that using similar font files and text input complement each other.

When asked, \textit{``Do you think you were able to effectively use the \textsc{Update all} button?''}, nine participants responded positively, with eight participants noting that it was more convenient than designing each character individually.
P10, who answered no, expressed dissatisfaction, commenting that it would be more convenient if users could toggle between adjusting either all characters at once or individually. 
In particular, he felt that having a feature to switch to individual adjustments is crucial during the fine-tuning stage.
We focus on the simplicity of the UI in this user study and this individual adjustments function is effective especially when designing many characters like all Roman characters.

In response to the question, ``\textit{Do you feel that you could design characters with a sense of agency using our system?}'', posed only to P5--P10, all six participants responded affirmatively.
P7 and P10 noted that in the baseline system, the line search space was initialized randomly, making the process feel highly dependent on luck. 
In contrast, they appreciated that our proposed system allowed them to control the initialization by specifying their preferences through multimodal references.
This insight aligns with the findings in \autoref{sec:simulatedEvaluation}, which show the initialization using multimodal references leads to better results compared to random initialization.
Additionally, P9 commented that he felt he could convey his intentions to the system by inputting texts.
These insights indicate that our system, leveraging multimodal references, provides users with a greater sense of agency compared to the baseline system.

Seven participants highlighted the usefulness of the history interface during the design process.
P5 remarked that the feature was particularly effective, as there were times when he felt a previous font was better.
In such cases, the history interface allowed him to revisit and continue from that point, saving effort.
He also noted the inconvenience of the baseline system lacking this feature.
P6 commented that comparing the current font displayed on the slider with previously created fonts helped him determine which one aligned more closely with his intended design.
He also said that in the baseline system, he found it challenging to reset after creating a satisfactory font.
In contrast, our system's history interface made him feel more confident about updating or resetting, as it allowed him to aim for even better results without hesitation.
This exemplifies that the history interface is useful not only for storing the designed characters and enabling the users to go back to a past point but also for making them advance the design process as boldly as they want.
It also indicates that the history interface reduces stress and increases freedom and creativity in the design task.

Overall, the feedback suggests that the proposed functions in our system effectively support font design for different participants based on their design preferences and familiarity with typography.

\section{Application Demonstrations}
\label{sec:demonstration}
In this section, we will show a case study where participants are required to design characters in more realistic situations and the adaptation to another writing system rather than Roman characters.

\subsection{Designing Fonts for Graphic Design Purposes}
\label{sec:demonstration_graphic_design}
In practical usage, it is important to evaluate whether our system can support font design for graphic design purposes, such as logo design or advertisement design, as noted by professional designers in \autoref{sec:limitations}.
To explore this, we asked participants to create suitable characters for specific design contexts.

To begin with the conclusion, from the feedback, we observed that participants using our system did not initially have a clear vision of the font they wanted. 
However, as they explored different font styles, they drew inspiration from the designs they encountered, ultimately creating their own unique characters.
While participants occasionally struggled with fine adjustments, such as correcting distorted lines, they generally felt they were able to create fonts that aligned with their intended concepts.
In the following sections, we present two design scenarios: conference logo design and advertisement poster design.
In the conference logo task, four participants (P11--P14) created characters for a logo, demonstrating a variety of font styles using our system.
In the advertisement poster task, another six participants (P15--P20) designed characters for different posters, tailoring their fonts to the target concepts.

\subsubsection{Design a Conference Logo}
\begin{figure}[ht]
    \centering
    \includegraphics[width=0.95\linewidth]{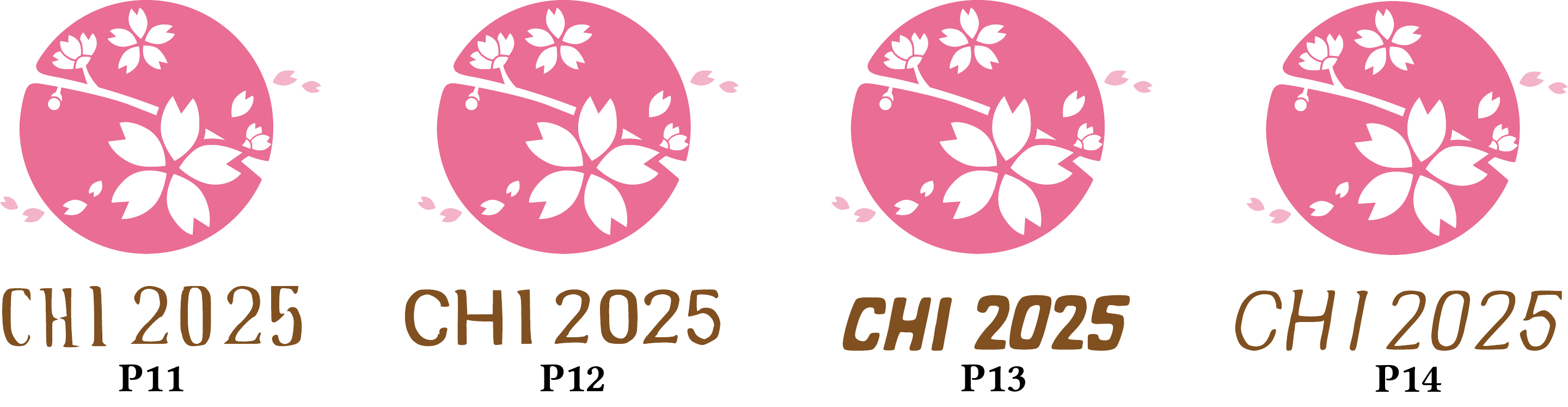}
    \caption{
    \textbf{Designed characters for the conference logo.}
    Four participants designed the characters for the conference logo.
    They designed a diverse range of fonts based on their unique sensibilities.
}
\label{fig:CHILogoDesign}
\end{figure}

In this experiment, participants with no prior font design experience were tasked with designing a conference logo.
Specifically, they were asked to create the characters ``CHI 2025'' to complement the cherry blossom motif in the conference logo.
After receiving an introduction to using our system, participants completed the task, and their feedback was collected.
During the design process, participants were only shown the cherry blossom logo and were not aware of the characters in the official conference logo.
As shown in \autoref{fig:CHILogoDesign}, the designs varied among participants.
P11 noted that her designed characters complemented the cherry blossom logo, highlighting that her favorite aspect was the fading central lines in ``H'' and ``5.''
This fading part appeared accidentally but complements the logo from her point of view, so she adopted it.
She also attempted to replicate this effect in ``2,'' but it was unsuccessful.
P12 said that he thought a decent and calm font was suitable for the conference logo and tried to make such a font.
He also commented that the font he designed was $90$ out of $100$ in terms of satisfaction, though his attempt to make ``I'' more straight was not successful.
P13 commented that he aimed to create a cute font inspired by the Japanese subculture, opting for a bold and rounded design.
For the initial step, he input the text, ``I want a cute and thick font" and found that the system performed as he hoped.
He also noted that the slider was effective in fine-tuning character details and eliminating unwanted distortions.
He was proud of the font he designed and believed it could be used in real-world applications, as the style of each character was well aligned.
P14 noted that he thought a thin and brush-style font fitted the Japanese-style logo and tried to make it.
He found multimodal input effective for the early stages of rough font design but felt it was less suitable for detailed exploration, ultimately relying on slider manipulation to refine the characters.
He was confident with the quality except for the noise and distortion in the designed characters.

\subsubsection{Design Advertisement Posters}
\begin{figure}[t]
    \centering
    \includegraphics[width=\linewidth]{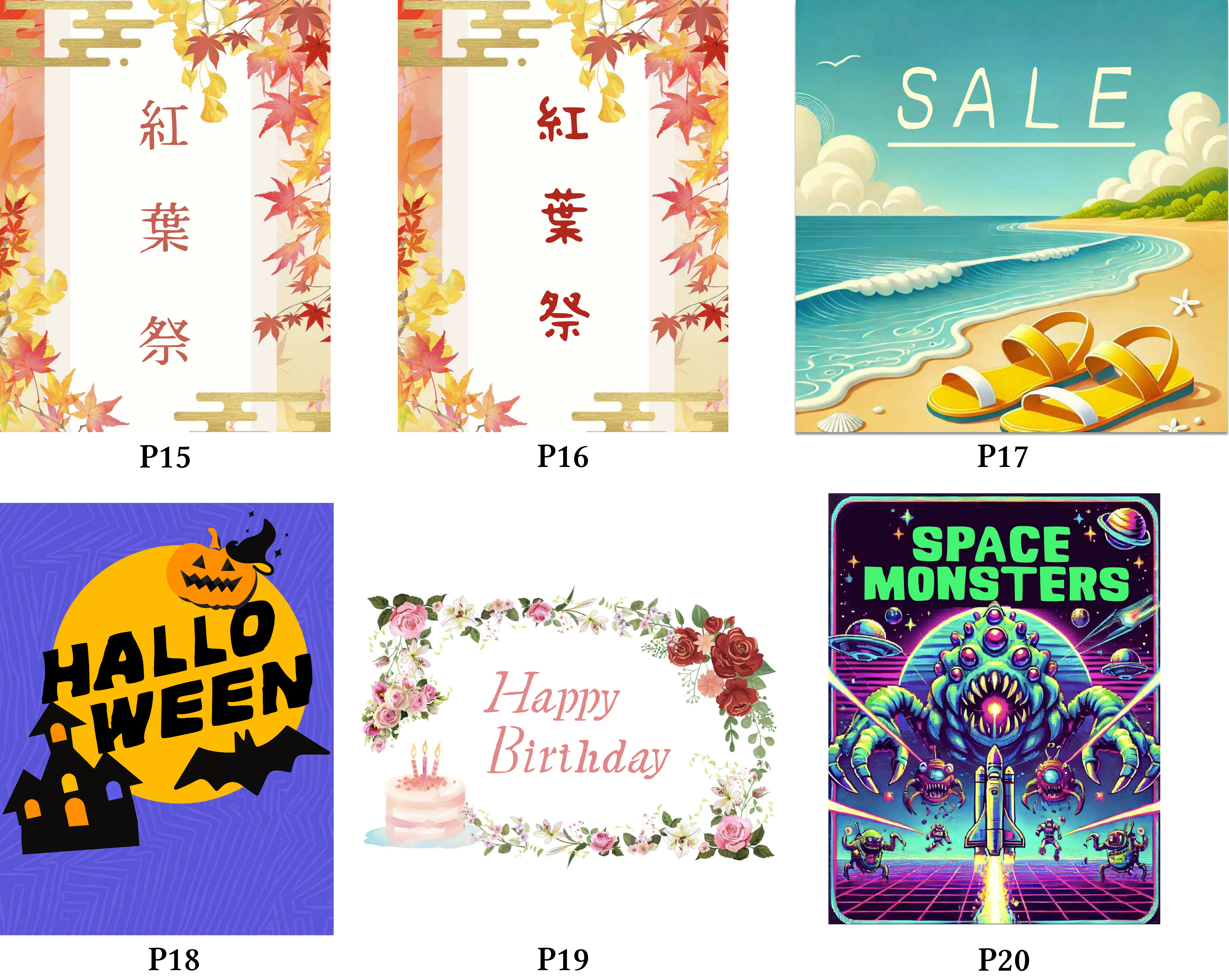}
    \caption{
    \textbf{Designed characters for the advertisement posters.}
    The participants designed the characters while viewing background images for the posters.
    The two posters on the top left are for an autumn foliage festival.
    P17 created a poster for a summer sale, while P18 designed characters for a Halloween event.
    P19 developed a font for a birthday card, and P20 created one for a movie poster.
}
\label{fig:poster}
\end{figure}

This demonstration shows font design for advertisement posters using our system, as illustrated in \autoref{fig:poster}.
During the design process, participants (P15--P20), who were introduced to the use of our system, were given a scenario and shown only the background image, with the task of creating characters that matched the visual context.
P15 and P16, both familiar with CJK writing systems, designed characters for an autumn foliage festival.
P5 rated his design $9$ out of $10$, expressing satisfaction with the traditional and formal font style he aimed to achieve.
He efficiently initialized the search space by inputting the text ``yu-mincho, serif'' (with ``yu-mincho'' being one of the most popular CJK fonts).
P16 commented that he envisioned a calm and warm font for the festival and was pleased with the result. 
He noted that their initial idea was simply based on the keyword ``warm,'' which he input into the system.
As he explored various styles, he gradually refined his design and reached a point of satisfaction.
P17, who designed the summer sale poster, aimed for a thin and refreshing font.
He observed elements in the background image such as the central white line, the seagull's wings, and the wave’s border, and decided that the character weight should align with these features.
By adjusting the slider, he was able to find a suitable font weight, though he expressed some dissatisfaction with the distortion of the top horizontal bar in the letter ``E.''

P18, tasked with designing a Halloween poster, felt that a twisted font suited the Halloween theme. 
She also believed a cute, handwritten style complemented the surrounding elements like the pumpkin, house, and bat, and was satisfied with the bold characters she created. 
She began by inputting a bold and italic font into the system, then continued refining the design using only the slider suggested by the Bayesian optimization process.
She expressed confidence in using the system, despite having no prior font design experience, and enjoyed the process. 
She effectively utilized the system's features, such as reverting to previous iterations via the history area when her exploration veered in an undesired direction, and she repeatedly refined each character after the initial style propagation.
For details on P18's design process, refer to the supplemental material.
P19 designed characters for a birthday card, aiming for cursive and fashionable font, and expressed satisfaction with the result.
P20 created a font for a movie poster, aiming for characters that were "scary," "thick," and "retro," with shapes fitting within a rectangular form (e.g., the shape of "S" resembling a rectangle).
While he was generally satisfied with the overall design, he found it challenging to achieve symmetry in characters like "A," "M," and "T."
Overall, although participants encountered challenges in addressing minor distortions and style inconsistencies, all expressed satisfaction with their final designs.

\subsection{Designing CJK Fonts}
\begin{figure}[b]
    \centering
    \includegraphics[width=\linewidth]{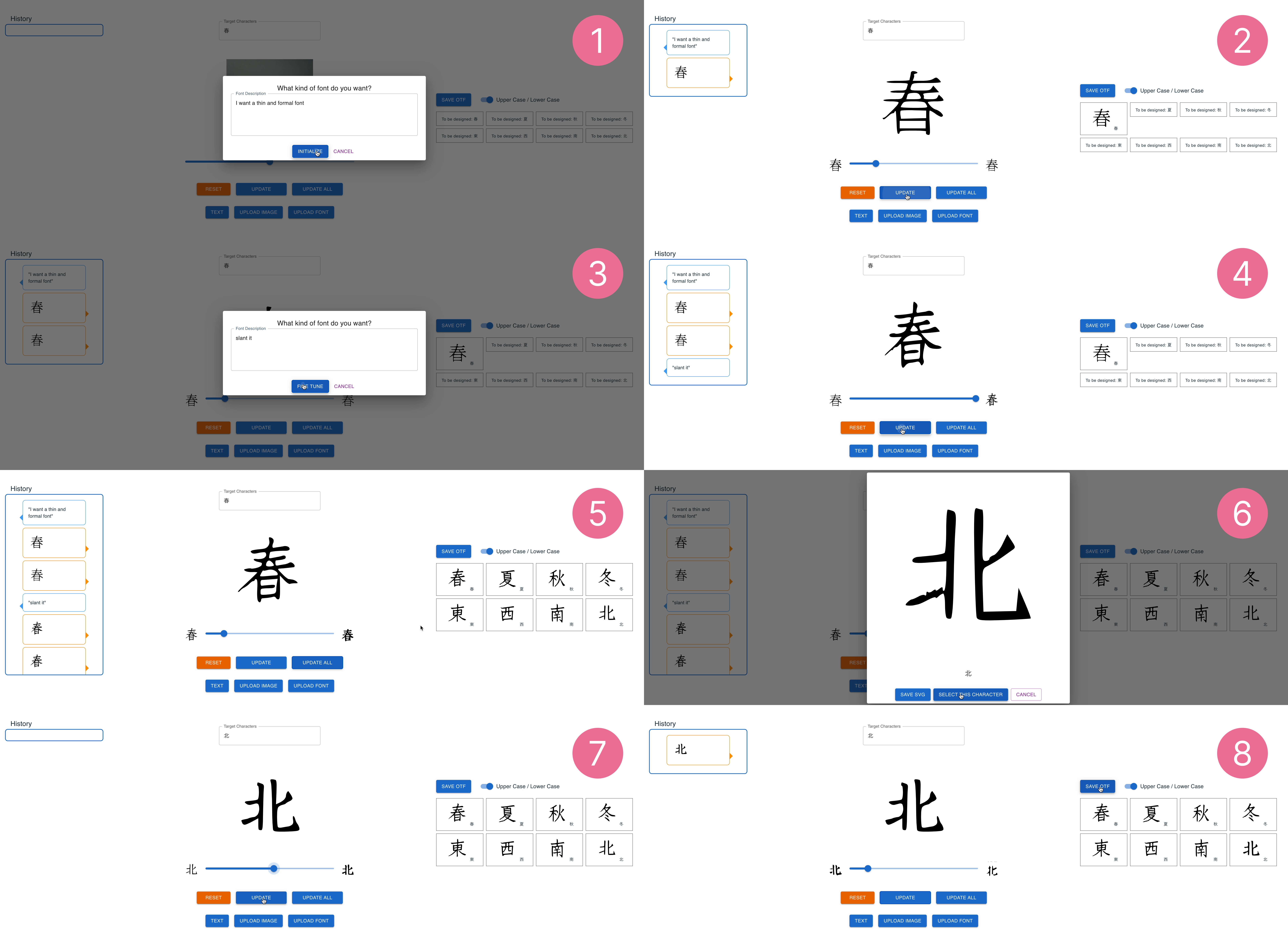}
    \caption{
    \textbf{Screenshots of CJK font character design.}
    The demonstration of CJK character designs using our system.
    The order is displayed in the upper right corner of each screenshot.
    See the supplemental material for the video.
}
\label{fig:CJKDesign}
\end{figure}

In the user study (\autoref{sec:user-study}), we demonstrated that participants could efficiently design Roman characters using our proposed system.
 By swapping the font generative model, the system can also support other writing systems, including Chinese, Japanese, and Korean (CJK).
As shown in \autoref{fig:CJKDesign}, users can efficiently design CJK characters without the need of predesigned examples.
Once the design process is complete, users can download their custom fonts as OTF files.

\section{Discussion, Limitations and Future Work}
\label{sec:limitations}

\paragraph{Professional Designers Interview}
In our user study and demonstration, we focused on non-expert users, as our system is designed to help them create fonts without requiring specialized knowledge.
However, feedback from professional designers (D1 and D2) is also crucial for identifying areas for improvement and enhancing the versatility of \systemName.
To gather feedback, we interviewed two professional designers from an advertising company, demonstrating our system and its outputs while asking two main questions:
1. \textit{"Is our system practical for designing advertisements in real-world scenarios?"}
2. \textit{"Are there any areas in our method that could be improved or features that should be added?"}

For the first question, both designers said they often need to create or customize unique characters for advertisement posters and logos.
They acknowledged that our system is suitable for such tasks, validating its use in designing characters for posters and conference logos.
For the second question, they pointed out that the distortion in the generated characters should be fixed for use in real-world scenarios.
Additionally, they expressed that a function allowing users to edit the generated characters is desirable.
Moreover, D1 noted that, for logo design, she sometimes needs to create each unique character in different styles; thus, the style propagation feature is not needed in such cases.
This feedback suggests that offering an option to disable style propagation could enhance user flexibility.
D2 mentioned that it would be beneficial to save all generated characters, visualize them in the UI, and allow users to revisit any previous points.
This implies that developing a more comprehensive history visualization feature could be a potential future work.

\paragraph{Limited Quality of the Generate Characters}
As noted by several participants in \autoref{sec:user-study} and \autoref{sec:demonstration}, generated characters exhibit distortions and visual artifacts.
Additionally, despite our system's ability to help users design style-consistent characters, some style inconsistencies persist.
For instance, during the demonstration of conference logo design in \autoref{sec:demonstration}, P11 mentioned the difficulty in applying the fading effect seen in ``H'' and ``5'' to ``2''.
These issues stem from the limitations of the font generative model~\cite{XieDGFont2021} and the vectorization method~\cite{selinger2003potrace} used in our system.
Our system requires a font generative model that can generate characters rapidly while manipulating the slider.
Therefore, even though some existing font generative models~\cite{liu2023dualvector,fu2024MSD, he2024difffont, liu2024qtfont,thamizharasan2024vecfusion} can produce characters with higher-quality and in more consistent styles, we cannot use them since they take longer time to generate.
For example, \textit{VecFusion}~\cite{thamizharasan2024vecfusion} takes 10 seconds to generate one glyph using A100, which is not feasible for interactive applications.
Meanwhile, this limitation also makes it harder for users to use our system to design style-consistent fonts containing many characters (\eg~$52$).
However, we believe that with the progress of font generative models, our system can utilize the latest models and generate fonts in higher quality.
To tackle these issues, we plan to explore how to optimize both the speed and quality of font generation. 
This includes integrating more advanced vectorization techniques or improving the efficiency of recent generative models without sacrificing real-time performance.

\paragraph{Broader Search Subspace for Multimodal References}
The current \textit{multimodal-guided subspace} is constructed by directly connecting a single encoded multimodal reference and previous user preference.
Although this approach enables users to explore the latent space around the multimodal reference, it may restrict other font style variations that are similar to the multimodal reference.
In the future, we plan to investigate how to construct a search subspace by connecting one or multiple multimodal references to an area in the style latent space that contains potential samples matching the desired styles similar to \cite{KoyamaGallery2020}.

\paragraph{Direct Character Editing}
As described in \autoref{sec:demonstration}, some users had difficulty creating desired characters due to character detail artifacts introduced by the font generative model and vectorization process.
In the future, to mitigate these artifacts, we plan to introduce painting tools that will allow users to edit the generated characters directly.

\paragraph{UI for Additional Typographical Support}
To design a font ready for production, it is crucial to include features that align characters with baseline, mean line, and other typographical guidelines including kerning.
These are essential for improving the overall appearance, alignment, and readability of the text.
We next plan to investigate how to integrate different user interfaces to facilitate font design that considers these typographical guidelines.

\section{Conclusion}
In this paper, we introduced \systemName, a new font design system that enables non-expert users to create fonts of any writing system without the need for predesigned characters.
Our system makes two main technical contributions: multimodal-guided subspace and retractable preference modeling, which address the two key limitations in existing human-in-the-loop PBO.
Additionally, we incorporated an iterative style propagation and refinement process, enabling users to design style consistent font.
Through a study, we demonstrated that non-expert users can efficiently design Roman characters using our system, supported by both quantitative and qualitative analysis.
Furthermore, we showcased how users could create Roman and CJK characters in realistic scenarios and achieve satisfying results.
\systemName is independent of any specific font generative model, making it adaptable to various models beyond \textit{DG-Font} used in this paper.
This flexibility ensures \systemName can evolve with future font generation technology advancements.

\begin{acks}
We thank the anonymous reviewers for their valuable feedback.
This work was partially supported by JST AdCORP, Grant Number JPMJKB2302, JSPS Grant-in-Aid JP23K16921, Japan, ANR-21-CE33-0002 GLACIS, France, and a collaboration with Dentsu Digital.
\end{acks}

\bibliographystyle{ACM-Reference-Format}
\bibliography{bib_font}


\begin{thebibliography}{44}

%%% the Web address.

\ifx \showCODEN    \undefined \def \showCODEN     #1{\unskip}     \fi
\ifx \showDOI      \undefined \def \showDOI       #1{#1}\fi
\ifx \showISBNx    \undefined \def \showISBNx     #1{\unskip}     \fi
\ifx \showISBNxiii \undefined \def \showISBNxiii  #1{\unskip}     \fi
\ifx \showISSN     \undefined \def \showISSN      #1{\unskip}     \fi
\ifx \showLCCN     \undefined \def \showLCCN      #1{\unskip}     \fi
\ifx \shownote     \undefined \def \shownote      #1{#1}          \fi
\ifx \showarticletitle \undefined \def \showarticletitle #1{#1}   \fi
\ifx \showURL      \undefined \def \showURL       {\relax}        \fi
% invisible to TeX
\providecommand\bibfield[2]{#2}
\providecommand\bibinfo[2]{#2}
\providecommand\natexlab[1]{#1}
\providecommand\showeprint[2][]{arXiv:#2}

\bibitem[Brochu et~al\mbox{.}(2010a)]%
        {Brochu2010A}
\bibfield{author}{\bibinfo{person}{Eric Brochu}, \bibinfo{person}{Tyson Brochu}, {and} \bibinfo{person}{Nando De~Freitas}.} \bibinfo{year}{2010}\natexlab{a}.
\newblock \showarticletitle{A Bayesian interactive optimization approach to procedural animation design}. In \bibinfo{booktitle}{\emph{Proceedings of the 2010 ACM SIGGRAPH/Eurographics Symposium on Computer Animation}}. \bibinfo{pages}{103--112}.
\newblock


\bibitem[Brochu et~al\mbox{.}(2010b)]%
        {Brochu2010B}
\bibfield{author}{\bibinfo{person}{Eric Brochu}, \bibinfo{person}{Vlad~M Cora}, {and} \bibinfo{person}{Nando De~Freitas}.} \bibinfo{year}{2010}\natexlab{b}.
\newblock \showarticletitle{A tutorial on Bayesian optimization of expensive cost functions, with application to active user modeling and hierarchical reinforcement learning}.
\newblock \bibinfo{journal}{\emph{arXiv preprint arXiv:1012.2599}} (\bibinfo{year}{2010}).
\newblock


\bibitem[Brochu et~al\mbox{.}(2007)]%
        {Brochu2007}
\bibfield{author}{\bibinfo{person}{Eric Brochu}, \bibinfo{person}{Nando~de Freitas}, {and} \bibinfo{person}{Abhijeet Ghosh}.} \bibinfo{year}{2007}\natexlab{}.
\newblock \showarticletitle{Active preference learning with discrete choice data}.
\newblock \bibinfo{journal}{\emph{Advances in neural information processing systems}}  \bibinfo{volume}{20} (\bibinfo{year}{2007}).
\newblock


\bibitem[Campbell and Kautz(2014)]%
        {CampbellFontManifold2014}
\bibfield{author}{\bibinfo{person}{Neill~DF Campbell} {and} \bibinfo{person}{Jan Kautz}.} \bibinfo{year}{2014}\natexlab{}.
\newblock \showarticletitle{Learning a manifold of fonts}.
\newblock \bibinfo{journal}{\emph{ACM Transactions on Graphics (ToG)}} \bibinfo{volume}{33}, \bibinfo{number}{4} (\bibinfo{year}{2014}), \bibinfo{pages}{1--11}.
\newblock


\bibitem[Cha et~al\mbox{.}(2020)]%
        {ChaDMFont2020}
\bibfield{author}{\bibinfo{person}{Junbum Cha}, \bibinfo{person}{Sanghyuk Chun}, \bibinfo{person}{Gayoung Lee}, \bibinfo{person}{Bado Lee}, \bibinfo{person}{Seonghyeon Kim}, {and} \bibinfo{person}{Hwalsuk Lee}.} \bibinfo{year}{2020}\natexlab{}.
\newblock \showarticletitle{Few-shot compositional font generation with dual memory}. In \bibinfo{booktitle}{\emph{European Conference on Computer Vision}}. Springer, \bibinfo{pages}{735--751}.
\newblock


\bibitem[Chan et~al\mbox{.}(2022)]%
        {Chan2022}
\bibfield{author}{\bibinfo{person}{Liwei Chan}, \bibinfo{person}{Yi-Chi Liao}, \bibinfo{person}{George~B Mo}, \bibinfo{person}{John~J Dudley}, \bibinfo{person}{Chun-Lien Cheng}, \bibinfo{person}{Per~Ola Kristensson}, {and} \bibinfo{person}{Antti Oulasvirta}.} \bibinfo{year}{2022}\natexlab{}.
\newblock \showarticletitle{Investigating Positive and Negative Qualities of Human-in-the-Loop Optimization for Designing Interaction Techniques}. In \bibinfo{booktitle}{\emph{Proceedings of the 2022 CHI Conference on Human Factors in Computing Systems}} \emph{(\bibinfo{series}{CHI '22})}. Article \bibinfo{articleno}{112}.
\newblock
\urldef\tempurl%
\url{https://doi.org/10.1145/3491102.3501850}
\showDOI{\tempurl}


\bibitem[Chong et~al\mbox{.}(2021)]%
        {chong2021interactive}
\bibfield{author}{\bibinfo{person}{Toby Chong}, \bibinfo{person}{I-Chao Shen}, \bibinfo{person}{Issei Sato}, {and} \bibinfo{person}{Takeo Igarashi}.} \bibinfo{year}{2021}\natexlab{}.
\newblock \showarticletitle{Interactive Optimization of Generative Image Modelling using Sequential Subspace Search and Content-based Guidance}. In \bibinfo{booktitle}{\emph{Computer Graphics Forum}}, Vol.~\bibinfo{volume}{40}. Wiley Online Library, \bibinfo{pages}{279--292}.
\newblock


\bibitem[Fu et~al\mbox{.}(2024)]%
        {fu2024MSD}
\bibfield{author}{\bibinfo{person}{Bin Fu}, \bibinfo{person}{Fanghua Yu}, \bibinfo{person}{Anran Liu}, \bibinfo{person}{Zixuan Wang}, \bibinfo{person}{Jie Wen}, \bibinfo{person}{Junjun He}, {and} \bibinfo{person}{Yu Qiao}.} \bibinfo{year}{2024}\natexlab{}.
\newblock \showarticletitle{Generate Like Experts: Multi-Stage Font Generation by Incorporating Font Transfer Process into Diffusion Models}. In \bibinfo{booktitle}{\emph{Proceedings of the IEEE/CVF Conference on Computer Vision and Pattern Recognition}}. \bibinfo{pages}{6892--6901}.
\newblock


\bibitem[Fu et~al\mbox{.}(2023)]%
        {fu2023dreamsim}
\bibfield{author}{\bibinfo{person}{Stephanie Fu}, \bibinfo{person}{Netanel~Y Tamir}, \bibinfo{person}{Shobhita Sundaram}, \bibinfo{person}{Lucy Chai}, \bibinfo{person}{Richard Zhang}, \bibinfo{person}{Tali Dekel}, {and} \bibinfo{person}{Phillip Isola}.} \bibinfo{year}{2023}\natexlab{}.
\newblock \showarticletitle{DreamSim: learning new dimensions of human visual similarity using synthetic data}. In \bibinfo{booktitle}{\emph{Proceedings of the 37th International Conference on Neural Information Processing Systems}}. \bibinfo{pages}{50742--50768}.
\newblock


\bibitem[He et~al\mbox{.}(2024)]%
        {he2024difffont}
\bibfield{author}{\bibinfo{person}{Haibin He}, \bibinfo{person}{Xinyuan Chen}, \bibinfo{person}{Chaoyue Wang}, \bibinfo{person}{Juhua Liu}, \bibinfo{person}{Bo Du}, \bibinfo{person}{Dacheng Tao}, {and} \bibinfo{person}{Qiao Yu}.} \bibinfo{year}{2024}\natexlab{}.
\newblock \showarticletitle{Diff-font: Diffusion model for robust one-shot font generation}.
\newblock \bibinfo{journal}{\emph{International Journal of Computer Vision}} (\bibinfo{year}{2024}), \bibinfo{pages}{1--15}.
\newblock


\bibitem[Ho et~al\mbox{.}(2020)]%
        {ho2020denoising}
\bibfield{author}{\bibinfo{person}{Jonathan Ho}, \bibinfo{person}{Ajay Jain}, {and} \bibinfo{person}{Pieter Abbeel}.} \bibinfo{year}{2020}\natexlab{}.
\newblock \showarticletitle{Denoising diffusion probabilistic models}.
\newblock \bibinfo{journal}{\emph{Advances in neural information processing systems}}  \bibinfo{volume}{33} (\bibinfo{year}{2020}), \bibinfo{pages}{6840--6851}.
\newblock


\bibitem[Huang and Belongie(2017)]%
        {HuangAdaIN2017}
\bibfield{author}{\bibinfo{person}{Xun Huang} {and} \bibinfo{person}{Serge Belongie}.} \bibinfo{year}{2017}\natexlab{}.
\newblock \showarticletitle{Arbitrary style transfer in real-time with adaptive instance normalization}. In \bibinfo{booktitle}{\emph{Proceedings of the IEEE international conference on computer vision}}. \bibinfo{pages}{1501--1510}.
\newblock


\bibitem[Jiang et~al\mbox{.}(2017)]%
        {JiangDCFont2017}
\bibfield{author}{\bibinfo{person}{Yue Jiang}, \bibinfo{person}{Zhouhui Lian}, \bibinfo{person}{Yingmin Tang}, {and} \bibinfo{person}{Jianguo Xiao}.} \bibinfo{year}{2017}\natexlab{}.
\newblock \showarticletitle{DCFont: an end-to-end deep Chinese font generation system}.
\newblock In \bibinfo{booktitle}{\emph{SIGGRAPH Asia 2017 Technical Briefs}}. \bibinfo{pages}{1--4}.
\newblock


\bibitem[Kadner et~al\mbox{.}(2021)]%
        {kadner2021adaptifont}
\bibfield{author}{\bibinfo{person}{Florian Kadner}, \bibinfo{person}{Yannik Keller}, {and} \bibinfo{person}{Constantin Rothkopf}.} \bibinfo{year}{2021}\natexlab{}.
\newblock \showarticletitle{Adaptifont: Increasing individuals’ reading speed with a generative font model and bayesian optimization}. In \bibinfo{booktitle}{\emph{Proceedings of the 2021 chi conference on human factors in computing systems}}. \bibinfo{pages}{1--11}.
\newblock


\bibitem[Koyama et~al\mbox{.}(2022)]%
        {Koyama2022}
\bibfield{author}{\bibinfo{person}{Yuki Koyama}, \bibinfo{person}{Toby Chong}, {and} \bibinfo{person}{Takeo Igarashi}.} \bibinfo{year}{2022}\natexlab{}.
\newblock \showarticletitle{Preferential Bayesian Optimisation for Visual Design}.
\newblock In \bibinfo{booktitle}{\emph{Bayesian Methods for Interaction and Design}}, \bibfield{editor}{\bibinfo{person}{John~H Williamson}, \bibinfo{person}{Antti Oulasvirta}, \bibinfo{person}{Per~Ola Kristensson}, {and} \bibinfo{person}{Nikola Banovic}} (Eds.). \bibinfo{publisher}{Cambridge University Press}, Chapter~8, \bibinfo{pages}{239--258}.
\newblock


\bibitem[Koyama et~al\mbox{.}(2020)]%
        {KoyamaGallery2020}
\bibfield{author}{\bibinfo{person}{Yuki Koyama}, \bibinfo{person}{Issei Sato}, {and} \bibinfo{person}{Masataka Goto}.} \bibinfo{year}{2020}\natexlab{}.
\newblock \showarticletitle{Sequential gallery for interactive visual design optimization}.
\newblock \bibinfo{journal}{\emph{ACM Transactions on Graphics (TOG)}} \bibinfo{volume}{39}, \bibinfo{number}{4} (\bibinfo{year}{2020}), \bibinfo{pages}{88--1}.
\newblock


\bibitem[Koyama et~al\mbox{.}(2017)]%
        {KoyamaSequential2017}
\bibfield{author}{\bibinfo{person}{Yuki Koyama}, \bibinfo{person}{Issei Sato}, \bibinfo{person}{Daisuke Sakamoto}, {and} \bibinfo{person}{Takeo Igarashi}.} \bibinfo{year}{2017}\natexlab{}.
\newblock \showarticletitle{Sequential line search for efficient visual design optimization by crowds}.
\newblock \bibinfo{journal}{\emph{ACM Transactions on Graphics (TOG)}} \bibinfo{volume}{36}, \bibinfo{number}{4} (\bibinfo{year}{2017}), \bibinfo{pages}{1--11}.
\newblock


\bibitem[Lian et~al\mbox{.}(2018)]%
        {LianEasyFont2017}
\bibfield{author}{\bibinfo{person}{Zhouhui Lian}, \bibinfo{person}{Bo Zhao}, \bibinfo{person}{Xudong Chen}, {and} \bibinfo{person}{Jianguo Xiao}.} \bibinfo{year}{2018}\natexlab{}.
\newblock \showarticletitle{EasyFont: a style learning-based system to easily build your large-scale handwriting fonts}.
\newblock \bibinfo{journal}{\emph{ACM Transactions on Graphics (TOG)}} \bibinfo{volume}{38}, \bibinfo{number}{1} (\bibinfo{year}{2018}), \bibinfo{pages}{1--18}.
\newblock


\bibitem[Liu and Lian(2024)]%
        {liu2024qtfont}
\bibfield{author}{\bibinfo{person}{Yitian Liu} {and} \bibinfo{person}{Zhouhui Lian}.} \bibinfo{year}{2024}\natexlab{}.
\newblock \showarticletitle{QT-Font: High-efficiency Font Synthesis via Quadtree-based Diffusion Models}. In \bibinfo{booktitle}{\emph{ACM SIGGRAPH 2024 Conference Papers}}. \bibinfo{pages}{1--11}.
\newblock


\bibitem[Liu et~al\mbox{.}(2023)]%
        {liu2023dualvector}
\bibfield{author}{\bibinfo{person}{Ying-Tian Liu}, \bibinfo{person}{Zhifei Zhang}, \bibinfo{person}{Yuan-Chen Guo}, \bibinfo{person}{Matthew Fisher}, \bibinfo{person}{Zhaowen Wang}, {and} \bibinfo{person}{Song-Hai Zhang}.} \bibinfo{year}{2023}\natexlab{}.
\newblock \showarticletitle{Dualvector: Unsupervised vector font synthesis with dual-part representation}. In \bibinfo{booktitle}{\emph{Proceedings of the IEEE/CVF Conference on Computer Vision and Pattern Recognition}}. \bibinfo{pages}{14193--14202}.
\newblock


\bibitem[Mo et~al\mbox{.}(2024)]%
        {Mo2024}
\bibfield{author}{\bibinfo{person}{George Mo}, \bibinfo{person}{John Dudley}, \bibinfo{person}{Liwei Chan}, \bibinfo{person}{Yi-Chi Liao}, \bibinfo{person}{Antti Oulasvirta}, {and} \bibinfo{person}{Per~Ola Kristensson}.} \bibinfo{year}{2024}\natexlab{}.
\newblock \showarticletitle{Cooperative Multi-Objective Bayesian Design Optimization}.
\newblock \bibinfo{journal}{\emph{ACM Trans. Interact. Intell. Syst.}} \bibinfo{volume}{14}, \bibinfo{number}{2} (\bibinfo{year}{2024}).
\newblock
\urldef\tempurl%
\url{https://doi.org/10.1145/3657643}
\showDOI{\tempurl}


\bibitem[O'Donovan et~al\mbox{.}(2014)]%
        {o2014exploratory}
\bibfield{author}{\bibinfo{person}{Peter O'Donovan}, \bibinfo{person}{Jundefinednis Lundefinedbeks}, \bibinfo{person}{Aseem Agarwala}, {and} \bibinfo{person}{Aaron Hertzmann}.} \bibinfo{year}{2014}\natexlab{}.
\newblock \showarticletitle{Exploratory Font Selection Using Crowdsourced Attributes}.
\newblock \bibinfo{journal}{\emph{ACM Transactions on Graphics}} \bibinfo{volume}{33}, \bibinfo{number}{4}, Article \bibinfo{articleno}{92} (\bibinfo{year}{2014}), \bibinfo{numpages}{9}~pages.
\newblock
\showISSN{0730-0301}
\urldef\tempurl%
\url{https://doi.org/10.1145/2601097.2601110}
\showDOI{\tempurl}


\bibitem[Park et~al\mbox{.}(2021)]%
        {ParkMultipleHeads2021}
\bibfield{author}{\bibinfo{person}{Song Park}, \bibinfo{person}{Sanghyuk Chun}, \bibinfo{person}{Junbum Cha}, \bibinfo{person}{Bado Lee}, {and} \bibinfo{person}{Hyunjung Shim}.} \bibinfo{year}{2021}\natexlab{}.
\newblock \showarticletitle{Multiple heads are better than one: Few-shot font generation with multiple localized experts}. In \bibinfo{booktitle}{\emph{Proceedings of the IEEE/CVF International Conference on Computer Vision}}. \bibinfo{pages}{13900--13909}.
\newblock


\bibitem[Selinger(2003)]%
        {selinger2003potrace}
\bibfield{author}{\bibinfo{person}{Peter Selinger}.} \bibinfo{year}{2003}\natexlab{}.
\newblock \bibinfo{title}{Potrace: a polygon-based tracing algorithm}.
\newblock
\newblock


\bibitem[Shahriari et~al\mbox{.}(2015)]%
        {ShahriaiReviewBO2016}
\bibfield{author}{\bibinfo{person}{Bobak Shahriari}, \bibinfo{person}{Kevin Swersky}, \bibinfo{person}{Ziyu Wang}, \bibinfo{person}{Ryan~P Adams}, {and} \bibinfo{person}{Nando De~Freitas}.} \bibinfo{year}{2015}\natexlab{}.
\newblock \showarticletitle{Taking the human out of the loop: A review of Bayesian optimization}.
\newblock \bibinfo{journal}{\emph{Proc. IEEE}} \bibinfo{volume}{104}, \bibinfo{number}{1} (\bibinfo{year}{2015}), \bibinfo{pages}{148--175}.
\newblock


\bibitem[Sun et~al\mbox{.}(2017)]%
        {SunSAVAE2018}
\bibfield{author}{\bibinfo{person}{Danyang Sun}, \bibinfo{person}{Tongzheng Ren}, \bibinfo{person}{Chongxun Li}, \bibinfo{person}{Hang Su}, {and} \bibinfo{person}{Jun Zhu}.} \bibinfo{year}{2017}\natexlab{}.
\newblock \showarticletitle{Learning to write stylized chinese characters by reading a handful of examples}.
\newblock \bibinfo{journal}{\emph{arXiv preprint arXiv:1712.06424}} (\bibinfo{year}{2017}).
\newblock


\bibitem[Suveeranont and Igarashi(2010)]%
        {Igarashi2010}
\bibfield{author}{\bibinfo{person}{Rapee Suveeranont} {and} \bibinfo{person}{Takeo Igarashi}.} \bibinfo{year}{2010}\natexlab{}.
\newblock \showarticletitle{Example-based automatic font generation}. In \bibinfo{booktitle}{\emph{International Symposium on Smart Graphics}}. Springer, \bibinfo{pages}{127--138}.
\newblock


\bibitem[Tatsukawa et~al\mbox{.}(2024)]%
        {tatsukawa2024fontclip}
\bibfield{author}{\bibinfo{person}{Yuki Tatsukawa}, \bibinfo{person}{I-Chao Shen}, \bibinfo{person}{Anran Qi}, \bibinfo{person}{Yuki Koyama}, \bibinfo{person}{Takeo Igarashi}, {and} \bibinfo{person}{Ariel Shamir}.} \bibinfo{year}{2024}\natexlab{}.
\newblock \showarticletitle{FontCLIP: A Semantic Typography Visual-Language Model for Multilingual Font Applications}.
\newblock \bibinfo{journal}{\emph{Computer Graphics Forum}} (\bibinfo{year}{2024}).
\newblock


\bibitem[Thamizharasan et~al\mbox{.}(2024)]%
        {thamizharasan2024vecfusion}
\bibfield{author}{\bibinfo{person}{Vikas Thamizharasan}, \bibinfo{person}{Difan Liu}, \bibinfo{person}{Shantanu Agarwal}, \bibinfo{person}{Matthew Fisher}, \bibinfo{person}{Micha{\"e}l Gharbi}, \bibinfo{person}{Oliver Wang}, \bibinfo{person}{Alec Jacobson}, {and} \bibinfo{person}{Evangelos Kalogerakis}.} \bibinfo{year}{2024}\natexlab{}.
\newblock \showarticletitle{VecFusion: Vector Font Generation with Diffusion}. In \bibinfo{booktitle}{\emph{Proceedings of the IEEE/CVF Conference on Computer Vision and Pattern Recognition}}. \bibinfo{pages}{7943--7952}.
\newblock


\bibitem[Tian.(2016)]%
        {YuchenRewrite2016}
\bibfield{author}{\bibinfo{person}{Yuchen Tian.}} \bibinfo{year}{2016}\natexlab{}.
\newblock \bibinfo{title}{Rewrite: Neural Style Transfer For Chinese Fonts}.  (\bibinfo{year}{2016}).
\newblock


\bibitem[Tian.(2017)]%
        {YuchenZi2zi2017}
\bibfield{author}{\bibinfo{person}{Yuchen Tian.}} \bibinfo{year}{2017}\natexlab{}.
\newblock \bibinfo{title}{zi2zi: Master Chinese Calligraphy with Conditional Adversarial Networks}.  (\bibinfo{year}{2017}).
\newblock


\bibitem[Upchurch et~al\mbox{.}(2016)]%
        {UpchurchA2Z2016}
\bibfield{author}{\bibinfo{person}{Paul Upchurch}, \bibinfo{person}{Noah Snavely}, {and} \bibinfo{person}{Kavita Bala}.} \bibinfo{year}{2016}\natexlab{}.
\newblock \showarticletitle{From A to Z: supervised transfer of style and content using deep neural network generators}.
\newblock \bibinfo{journal}{\emph{arXiv preprint arXiv:1603.02003}} (\bibinfo{year}{2016}).
\newblock


\bibitem[Wang et~al\mbox{.}(2020)]%
        {wang2020attribute2font}
\bibfield{author}{\bibinfo{person}{Yizhi Wang}, \bibinfo{person}{Yue Gao}, {and} \bibinfo{person}{Zhouhui Lian}.} \bibinfo{year}{2020}\natexlab{}.
\newblock \showarticletitle{Attribute2font: Creating fonts you want from attributes}.
\newblock \bibinfo{journal}{\emph{ACM Transactions on Graphics (TOG)}} \bibinfo{volume}{39}, \bibinfo{number}{4} (\bibinfo{year}{2020}), \bibinfo{pages}{69--1}.
\newblock


\bibitem[Wang and Lian(2021)]%
        {wang2021deepvecfont}
\bibfield{author}{\bibinfo{person}{Yizhi Wang} {and} \bibinfo{person}{Zhouhui Lian}.} \bibinfo{year}{2021}\natexlab{}.
\newblock \showarticletitle{Deepvecfont: synthesizing high-quality vector fonts via dual-modality learning}.
\newblock \bibinfo{journal}{\emph{ACM Transactions on Graphics (TOG)}} \bibinfo{volume}{40}, \bibinfo{number}{6} (\bibinfo{year}{2021}), \bibinfo{pages}{1--15}.
\newblock


\bibitem[Wang et~al\mbox{.}(2023)]%
        {wang2023deepvecfontv2}
\bibfield{author}{\bibinfo{person}{Yuqing Wang}, \bibinfo{person}{Yizhi Wang}, \bibinfo{person}{Longhui Yu}, \bibinfo{person}{Yuesheng Zhu}, {and} \bibinfo{person}{Zhouhui Lian}.} \bibinfo{year}{2023}\natexlab{}.
\newblock \showarticletitle{Deepvecfont-v2: Exploiting transformers to synthesize vector fonts with higher quality}. In \bibinfo{booktitle}{\emph{Proceedings of the IEEE/CVF Conference on Computer Vision and Pattern Recognition}}. \bibinfo{pages}{18320--18328}.
\newblock


\bibitem[Xia et~al\mbox{.}(2023)]%
        {xia2023vecfontsdf}
\bibfield{author}{\bibinfo{person}{Zeqing Xia}, \bibinfo{person}{Bojun Xiong}, {and} \bibinfo{person}{Zhouhui Lian}.} \bibinfo{year}{2023}\natexlab{}.
\newblock \showarticletitle{Vecfontsdf: Learning to reconstruct and synthesize high-quality vector fonts via signed distance functions}. In \bibinfo{booktitle}{\emph{Proceedings of the IEEE/CVF Conference on Computer Vision and Pattern Recognition}}. \bibinfo{pages}{1848--1857}.
\newblock


\bibitem[Xie et~al\mbox{.}(2021)]%
        {XieDGFont2021}
\bibfield{author}{\bibinfo{person}{Yangchen Xie}, \bibinfo{person}{Xinyuan Chen}, \bibinfo{person}{Li Sun}, {and} \bibinfo{person}{Yue Lu}.} \bibinfo{year}{2021}\natexlab{}.
\newblock \showarticletitle{Dg-font: Deformable generative networks for unsupervised font generation}. In \bibinfo{booktitle}{\emph{Proceedings of the IEEE/CVF Conference on Computer Vision and Pattern Recognition}}. \bibinfo{pages}{5130--5140}.
\newblock


\bibitem[Xu et~al\mbox{.}(2009)]%
        {SounghuaAutomaticGeneration2009}
\bibfield{author}{\bibinfo{person}{Songhua Xu}, \bibinfo{person}{Tao Jin}, \bibinfo{person}{Hao Jiang}, {and} \bibinfo{person}{Francis~CM Lau}.} \bibinfo{year}{2009}\natexlab{}.
\newblock \showarticletitle{Automatic generation of personal chinese handwriting by capturing the characteristics of personal handwriting}. In \bibinfo{booktitle}{\emph{Twenty-First IAAI Conference}}.
\newblock


\bibitem[Yamamoto et~al\mbox{.}(2022)]%
        {bayelight}
\bibfield{author}{\bibinfo{person}{Kenta Yamamoto}, \bibinfo{person}{Yuki Koyama}, {and} \bibinfo{person}{Yoichi Ochiai}.} \bibinfo{year}{2022}\natexlab{}.
\newblock \showarticletitle{Photographic Lighting Design with Photographer-in-the-Loop Bayesian Optimization} \emph{(\bibinfo{series}{UIST '22})}. \bibinfo{publisher}{Association for Computing Machinery}, \bibinfo{address}{New York, NY, USA}, Article \bibinfo{articleno}{92}, \bibinfo{numpages}{11}~pages.
\newblock
\showISBNx{9781450393201}
\urldef\tempurl%
\url{https://doi.org/10.1145/3526113.3545690}
\showDOI{\tempurl}


\bibitem[Yang et~al\mbox{.}(2024)]%
        {yang2024fontdiffuser}
\bibfield{author}{\bibinfo{person}{Zhenhua Yang}, \bibinfo{person}{Dezhi Peng}, \bibinfo{person}{Yuxin Kong}, \bibinfo{person}{Yuyi Zhang}, \bibinfo{person}{Cong Yao}, {and} \bibinfo{person}{Lianwen Jin}.} \bibinfo{year}{2024}\natexlab{}.
\newblock \showarticletitle{Fontdiffuser: One-shot font generation via denoising diffusion with multi-scale content aggregation and style contrastive learning}. In \bibinfo{booktitle}{\emph{Proceedings of the AAAI conference on artificial intelligence}}, Vol.~\bibinfo{volume}{38}. \bibinfo{pages}{6603--6611}.
\newblock


\bibitem[Zhang et~al\mbox{.}(2018)]%
        {ZhangEMD2018}
\bibfield{author}{\bibinfo{person}{Yexun Zhang}, \bibinfo{person}{Ya Zhang}, {and} \bibinfo{person}{Wenbin Cai}.} \bibinfo{year}{2018}\natexlab{}.
\newblock \showarticletitle{Separating Style and Content for Generalized Style Transfer}. In \bibinfo{booktitle}{\emph{2018 IEEE/CVF Conference on Computer Vision and Pattern Recognition}}. \bibinfo{pages}{8447--8455}.
\newblock


\bibitem[Zhou et~al\mbox{.}(2011)]%
        {ZhouEasyGeneration2011}
\bibfield{author}{\bibinfo{person}{Baoyao Zhou}, \bibinfo{person}{Weihong Wang}, {and} \bibinfo{person}{Zhanghui Chen}.} \bibinfo{year}{2011}\natexlab{}.
\newblock \showarticletitle{Easy generation of personal Chinese handwritten fonts}. In \bibinfo{booktitle}{\emph{2011 IEEE international conference on multimedia and expo}}. IEEE, \bibinfo{pages}{1--6}.
\newblock


\bibitem[Zhou et~al\mbox{.}(2020)]%
        {ZhouGenerativeMelody2020}
\bibfield{author}{\bibinfo{person}{Yijun Zhou}, \bibinfo{person}{Yuki Koyama}, \bibinfo{person}{Masataka Goto}, {and} \bibinfo{person}{Takeo Igarashi}.} \bibinfo{year}{2020}\natexlab{}.
\newblock \showarticletitle{Generative melody composition with human-in-the-loop bayesian optimization}.
\newblock \bibinfo{journal}{\emph{arXiv preprint arXiv:2010.03190}} (\bibinfo{year}{2020}).
\newblock


\bibitem[Zong and Zhu(2014)]%
        {ZongStrokeBank2014}
\bibfield{author}{\bibinfo{person}{Alfred Zong} {and} \bibinfo{person}{Yuke Zhu}.} \bibinfo{year}{2014}\natexlab{}.
\newblock \showarticletitle{Strokebank: Automating personalized chinese handwriting generation}. In \bibinfo{booktitle}{\emph{Twenty-Sixth IAAI Conference}}.
\newblock


\end{thebibliography}

\end{document}